\PassOptionsToPackage{table}{xcolor}
\documentclass[acmsmall]{acmart}

\setcopyright{cc}
\setcctype{by}
\acmJournal{PACMSE}
\acmYear{2026} \acmVolume{3} \acmNumber{ISSTA} \acmArticle{ISSTA029}
\acmMonth{10} \acmDOI{10.1145/3832120}

\AtBeginDocument{
  }

\usepackage[utf8]{inputenc}
\usepackage[T1]{fontenc}
\usepackage{hyperref}
\usepackage{url}
\usepackage{booktabs}
\usepackage{amsfonts}
\usepackage{nicefrac}
\usepackage{microtype}
\usepackage[ruled,vlined]{algorithm2e}
\usepackage{amsmath}
\usepackage{xspace}
\usepackage{flushend}
\usepackage{multirow}
\usepackage{graphicx}
\usepackage{array}
\usepackage{fontawesome}
\usepackage{enumitem}
\usepackage{makecell}
\usepackage{threeparttable}
 \usepackage{pifont}

\usepackage{textcomp}
\usepackage{upquote}

\usepackage{lipsum}
\usepackage[most]{tcolorbox}

\newcommand{\name}{{SEER}\xspace}

\newcounter{finding}
\newcommand*{\newfinding}{Finding~\refstepcounter{finding}\thefinding:~}

\newtcolorbox{promptbox}{
  colback=blue!5!white,
  colframe=blue!30!black,
  fonttitle=\bfseries,
  title=Prompt,
  rounded corners,
  boxrule=1pt,
  sharp corners=south,
  halign=flush center,
}

\newtcolorbox{leftlinebox}{
  enhanced,
  boxrule=0pt,frame hidden,
  borderline west={4pt}{0pt}{gray!50!white},
  colback=gray!5!white,
  sharp corners,
  overlay={
    \node[anchor=north east] at (frame.north east) {\color{gray}\faPencil}; 
  }
}

\newif\ifhideTrace\hideTracefalse
\newcommand{\dx}[1]{\ifhideTrace#1\else\textcolor{magenta}{#1}\fi}

\hideTracetrue

\title[SEER: Self-Enhancing Chain-of-Thought Compression for Reasoning Models]{SEER: Self-Enhancing Chain-of-Thought Compression for Reasoning Models}

\author{Kerui Huang}
\orcid{0009-0005-7396-5027}
\affiliation{
  \department{The State Key Laboratory of Blockchain and Data Security}
  \institution{Zhejiang University}
  \city{Hangzhou}
  \state{Zhejiang}
  \country{China}
}
\email{huangkerui@zju.edu.cn}

\author{Shuhan Liu}
\orcid{0009-0000-2848-3735}
\affiliation{
  \department{The State Key Laboratory of Blockchain and Data Security}
  \institution{Zhejiang University}
  \city{Hangzhou}
  \state{Zhejiang}
  \country{China}
}
\email{liushuhan@zju.edu.cn}

\author{Xing Hu}
\orcid{0000-0003-0093-3292}
\affiliation{
  \department{The State Key Laboratory of Blockchain and Data Security}
  \institution{Zhejiang University}
  \city{Hangzhou}
  \state{Zhejiang}
  \country{China}
}
\email{xinghu@zju.edu.cn}

\author{Tongtong Xu}
\orcid{0000-0002-4323-497X}
\affiliation{
  \department{State Key Laboratory for Novel Software Technology}
  \institution{Nanjing University}
  \city{Nanjing}
  \state{Jiangsu}
  \country{China}
}
\email{xttluck@gmail.com}

\author{Lingfeng Bao}
\orcid{0000-0003-1846-0921}
\affiliation{
  \department{The State Key Laboratory of Blockchain and Data Security}
  \institution{Zhejiang University}
  \city{Hangzhou}
  \state{Zhejiang}
  \country{China}
}
\email{lingfengbao@zju.edu.cn}

\author{Xin Xia}
\orcid{0000-0002-6302-3256}
\authornote{Corresponding author.}
\affiliation{%
  \department{The State Key Laboratory of Blockchain and Data Security} 
  \institution{Zhejiang University}
  \city{Hangzhou}
  \state{Zhejiang}
  \country{China}
}
\affiliation{%
  \institution{Hangzhou High-Tech Zone (Binjiang) Institute of Blockchain and Data Security}
  \city{Hangzhou}
  \state{Zhejiang}
  \country{China}
}
\email{xin.xia@acm.org}

\begin{document}

\begin{abstract}
Chain-of-Thought (CoT) prompting can substantially improve the reasoning ability of large language models (LLMs), but it often comes with high inference cost due to long and poorly controlled reasoning traces.
This overhead is particularly problematic in software engineering tasks (e.g., code generation), where both latency and output reliability matter.

To better understand this trade-off, we conduct an empirical study on widely used code generation benchmarks and observe that many modern reasoning models produce excessively verbose CoTs (often thousands of tokens), which frequently leads to truncation and unstable generation.
Using a strict $n$-gram repetition detector, we find that  most observed truncations are associated with degenerate looping behaviors.  \dx{In addition, a HumanEval/129 case study shows that failed generations can be longer than successful ones, suggesting limited returns from overlong reasoning.}

Motivated by these findings, we propose \textbf{SEER} (\textbf{\underline{S}}elf-\textbf{\underline{E}}nhancing \textbf{\underline{E}}fficient \textbf{\underline{R}}easoning), a self-enhancing framework for adaptive CoT compression.
\name improves the conciseness of reasoning while preserving output quality, without relying on external compression tools.
\name refines self-generated CoT data via Best-of-N sampling to suppress looping and redundant traces, then applies a lightweight, data-driven filter to encourage concise yet correct reasoning.
It then fine-tunes the model on the filtered data to internalize concise reasoning behaviors.
Across  four software engineering benchmarks on the evaluated \texttt{DeepSeek-R1-Distill-Qwen-7B} backbone, \name reduces CoT length by  34.6\% on average while improving task performance,  \dx{with reduced truncation and fewer reasoning loops}.

\end{abstract}



\maketitle

\section{Introduction}
The introduction of Chain-of-Thought (CoT) marks a pivotal advancement in enhancing the reasoning capabilities of Large Language Models (LLMs), particularly for complex, multi-step tasks. 
By encouraging models to explicitly articulate intermediate reasoning steps, 
CoT enables a step-by-step breakdown of problems rather than producing an answer in a single attempt~\cite{NEURIPS2022_8bb0d291,wei2022chain}.
Existing studies have shown that CoT achieves notable improvements across tasks, including arithmetic, logic, and commonsense reasoning~\cite{nye2021workscratchpadsintermediatecomputation,wang2023selfconsistency,zhou2023leasttomost}.
Recently, the utilization of CoT has become more important in software engineering tasks, since they typically require precise logical reasoning, structured multi-step problem solving, and decision-making~\cite{yao2023tree,li2025structured}.

\begin{figure*}[ht]
  \centering
  \includegraphics[width=0.95\textwidth]{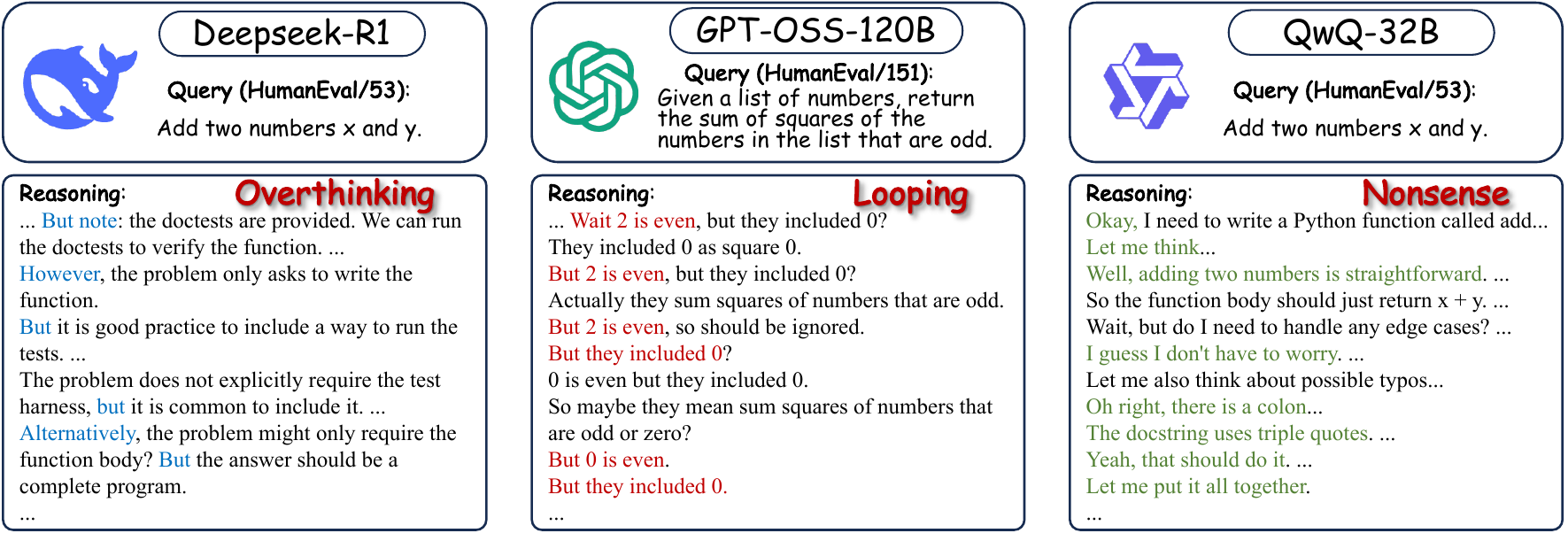} 
  \caption{Examples of Common Chain-of-Thought Issues Across Different Models} 
  \label{fig:example} 
\end{figure*}

Figure~\ref{fig:example} presents examples of common CoT issues observed across different open-source reasoning models.
Even state-of-the-art reasoning LLMs may generate excessive or poorly controlled CoT, including overthinking, repetitive reasoning, and semantically unproductive deliberation. This issue can arise even in simple tasks with concise solutions.
Such verbosity increases inference cost and latency, which is particularly problematic in code‑related scenarios.  
In software‑engineering contexts where reasoning is part of agent‑based workflows, uncontrolled CoT can increase system overhead and cause unnecessary reasoning steps, as observed in AutoGPT and LangGraph~\cite{autogpt_do_nothing,langgraph_repetition,autogpt_recursive_loop}.
A common case is that an agent repeatedly re‑examines the same action or planning step without advancing, which wastes tokens and slows execution. 

To examine the effect of verbose CoTs in real-world settings, we conduct a systematic empirical analysis of CoT reasoning across a diverse set of open-source models on code generation tasks.
We observe that excessively long CoT is common: most reasoning models generate 2K to 4K CoT tokens on average, and truncation can reach up to 52.7\% under a 16K token budget. This verbosity is closely associated with reduced efficiency and reliability, as enabling CoT can increase token cost by up to an order of magnitude. Using a strict $n$-gram repetition detector, we find that 2,848 out of 3,298 truncations (86.4\%) are identified as looping, and manual inspection suggests that this figure is a conservative lower bound. Our empirical results also show that prompt-based conciseness control is limited and highly model-dependent.
Overall, these findings highlight the need for effective mechanisms to regulate reasoning length and mitigate looping behaviors, i.e., CoT compression methods.

Prior work on CoT compression can be generally divided into two categories, one of which applies explicit compression operations to CoT, for example, TokenSkip with LLMLingua and C3oT with GPT-4~\cite{xia_tokenskip_2025,kang_c3ot_2025,zhuang2025accelerating}.
These methods usually achieve high compression ratios, but they often cause information loss and ``thought leap'' effects~\cite{xu2025mind}.
The second line of work relies on repeated sampling or distillation to collect simpler reasoning examples and encourage the model to produce more concise reasoning, such as Naive BoN~\cite{munkhbat2025selftraining,yu_distilling_2024}.
This form of implicit compression avoids information loss, but typically cannot achieve a high level of compression.

In this paper, we propose \textbf{SEER} (\textbf{\underline{S}}elf-\textbf{\underline{E}}nhancing \textbf{\underline{E}}fficient \textbf{\underline{R}}easoning), an adaptive framework designed to make CoT reasoning both effective and efficient for software engineering tasks.
Rather than relying on external compression tools, \name internalizes CoT control into the training process by learning concise reasoning patterns directly from self-generated outputs.
\name adopts a self-enhancing paradigm in which the model generates multiple CoT-augmented candidates and refines them through two complementary mechanisms.
First, Best-of-N (BoN) sampling explicitly addresses looping behaviors by selecting the shortest correct reasoning trace among multiple candidates, effectively suppressing loops and redundant expansions observed in our empirical study.
Second, \name introduces an adaptive CoT filtering strategy that calibrates maximum reasoning length based on dataset-specific length distributions. This mechanism is grounded in \dx{our empirical observation that useful reasoning traces in the evaluated settings are often much shorter than raw CoTs, helping prevent both over-compression and excessive verbosity}.

We evaluate \name on three representative software engineering tasks: code generation, defect detection and natural language code search.
Across the evaluated tasks on the \texttt{DeepSeek-R1-Distill- Qwen-7B} model, \name reduces CoT length by an average of 34.6\% while preserving or even improving pass@1 accuracy. Compared with the evaluated baselines, \name demonstrates \dx{the best overall trade-off}, achieving the highest pass@1 accuracy while also delivering the most substantial reduction in CoT length on code-related tasks.
Importantly, \name substantially mitigates truncation and infinite reasoning loops.
As a result, reasoning loops are mitigated by up to 96.8\%, leading to improved inference efficiency and output stability.
In summary, our main contributions are as follows:
\begin{itemize}[leftmargin=*]
    \item We present a systematic empirical study to analyze the impact of long CoT reasoning in code generation. Our results indicate that excessive CoT is common and truncation is an important factor affecting model performance. \dx{In the evaluated setting, most observed truncations are associated with reasoning loops, which lower accuracy and make generation unstable.}

    \item We propose \name, a lightweight and self-enhancing CoT compression framework that 
    makes CoT reasoning both effective and efficient for software engineering tasks.
    By learning concise reasoning patterns from self-generated outputs, \name effectively suppresses reasoning loops and truncation without relying on external compression tools or manual annotations.

    \item We evaluate \name across multiple software engineering tasks. Experimental results show that \name achieves an average \textbf{34.6\% reduction} in CoT length while preserving or even improving pass@1 accuracy on the evaluated backbone, alongside improved robustness against truncation and overlong reasoning.

\end{itemize}

\noindent\textbf{Paper Organization.} The remainder of this paper is structured as follows: Section~\ref{sec:background} reviews CoT efficiency challenges, the overthinking phenomenon, and self-generated fine-tuning. Section~\ref{sec:empirical} presents a motivating empirical study on the impact of CoT length. Section~\ref{sec:method} details the \name framework, including BoN sampling and adaptive filtering. Section~\ref{sec:setup} outlines the datasets, baselines, and evaluation metrics. Section~\ref{sec:evaluation} reports experimental findings and ablation results, addressing performance, generalizability, and loop mitigation. Section~\ref{sec:discussion} discusses broader implications and limitations, and Section~\ref{sec:related_work} reviews related literature. Finally, Section~\ref{sec:conclusion} concludes the paper and outlines future research directions.

\section{Background}\label{sec:background}

\subsection{The Efficiency of CoT}

LLMs have achieved remarkable success in complex reasoning tasks, largely due to advanced prompting techniques like CoT~\cite{wei2022chain}. CoT elicits multi-step reasoning by instructing the model to generate a series of intermediate steps before producing a final answer. This explicit articulation of the thought process allows the model to deconstruct complex problems, track its progress, and correct intermediate errors, leading to significant performance gains in tasks that demand logical deduction, arithmetic calculation, and symbolic reasoning.

However, the verbosity of CoT introduces significant computational overhead. Each reasoning step requires additional token generation, directly increasing inference latency and resource consumption, particularly for the KV-cache~\cite{agrawal2024taming,ho2024advances}. This overhead poses a critical bottleneck for real-time applications or systems with constrained computational budgets. Consequently, a key research direction in CoT efficiency is to strike an optimal balance: retaining the reasoning benefits of CoT while minimizing its associated costs. Approaches in this area typically explore methods such as pruning redundant steps, learning to generate more concise reasoning paths~\cite{kang_c3ot_2025, xia_tokenskip_2025}, or selectively applying CoT based on task complexity~\cite{wang2025think}.

\subsection{Existing Issues in CoT}
\noindent\faHandORight~\textbf{Overthinking Phenomenon in CoT.}
Beyond simple inefficiency, the relationship between CoT length and accuracy is not a straightforward ``longer is better.'' Prior work shows that longer reasoning does not always translate into higher accuracy~\cite{xie2025logic}. Instead, many studies identify an ``overthinking'' phenomenon. Accuracy often improves with additional reasoning at the early stages, since the model has more opportunity to explore possible solution paths. However, once the reasoning exceeds an optimal range, performance begins to decline as the model produces redundant or even erroneous content~\cite{chen2024not,jin2024impact,cuadron2025danger}. This decline is a direct result of overthinking, where excessive reasoning introduces noise, deviates from the correct reasoning path, and ultimately undermines accuracy~\cite{chen2024unlocking}. 
Overall, these findings highlight that the key challenge is to carefully control CoT length to reduce overthinking while preserving accuracy.

\noindent\faHandORight~\textbf{Looping Problem in CoT.}
Recent studies have revealed looping problems in CoT reasoning~\cite{pipis2025reasoningloop}, where LLMs repeatedly generate identical or semantically redundant reasoning segments instead of making progress toward a solution. 
This phenomenon also causes practical issues in real-world applications, as users have reported failures induced by looping behavior on various developer forums and agent frameworks~\cite{autogpt_do_nothing,langgraph_repetition,autogpt_recursive_loop}. 
Looping is particularly prominent under greedy or low-temperature decoding and has been observed across a wide range of modern reasoning models. 
Prior work suggests that this behavior stems from misestimated action probabilities during training: when advancing to the correct next reasoning step is difficult, models may favor repetitive continuations that are easier to predict, resulting in self-reinforcing loops. 
Such errors are further amplified by deterministic decoding, allowing repetition to persist across multiple steps. 
Although increasing sampling randomness can mitigate looping to some extent, it does not address the underlying learning deficiencies, indicating that looping is a fundamental inefficiency of CoT reasoning rather than a purely inference-time artifact.

\section{Empirical Study}\label{sec:empirical}
We first conduct an empirical study to investigate how the length of CoT reasoning influences the performance, efficiency, and stability of LLMs in software engineering scenarios. 
We choose a representative problem in software engineering, i.e., code generation tasks, as the subject of our empirical study.
This study provides a quantitative assessment of the challenges introduced by excessively long CoT outputs, thereby motivating the design of our method.

\subsection{Settings}
\subsubsection{Datasets}
Our empirical study is conducted on two widely used benchmarks, i.e.,  \textbf{HumanEval}~\cite{chen2021evaluating} and \textbf{MBPP-Sanitized}~\cite{austin2021program}.
They consist of Python programming problems, each accompanied by a function signature, a natural-language docstring describing the task, and a set of unit tests used to evaluate functional correctness.

\subsubsection{LLMs}
In our empirical study, we evaluate CoT behavior using a set of open-source reasoning-capable LLMs that are tested under the same experimental settings.
The models cover different parameter scales and architectures, including the \texttt{DeepSeek-R1-Distill-Qwen} family (\texttt{7B}, \texttt{14B}, and \texttt{32B}), \texttt{DeepSeek-R1-Distill-Llama-8B}, as well as other widely used reasoning models such as \texttt{QwQ-32B}, \texttt{Qwen3-8B}, \dx{\texttt{Qwen3.5-122B}, \texttt{Qwen3.5-397B},} and \texttt{gpt-oss-20b}.

\subsubsection{Evaluation Metrics}
To assess the performance of CoT on software engineering tasks, we employ three metrics.

\noindent\faAngleRight~\textbf{Pass@1.} Pass@1 calculates the proportion of problems for which the model produces a correct solution on the first attempt, directly reflecting its practical effectiveness in solving tasks. It is widely used on code generation tasks as it emphasizes single-attempt correctness under realistic deployment scenarios~\cite{chen2021evaluating}.

\noindent\faAngleRight~\textbf{Average Token Length.} We report the average number of tokens generated in model outputs. It reflects the verbosity of the reasoning process and helps assess whether excessively long outputs introduce redundancy or cause truncation. It also facilitates a direct comparison of output lengths among models with and without CoT reasoning.

\noindent\faAngleRight~\textbf{Truncation Number.} We count the number of instances where generation is terminated due to exceeding the maximum token budget. This metric reflects the risk of uncontrolled or redundant reasoning, which may prevent the model from completing valid solutions and thus degrade real-world usability.

\subsubsection{Implementation Details}~\label{sec:empirical_implement_detail} 
To ensure a fair and reliable empirical evaluation, we describe the implementation details of our experimental setup in this subsection.

\noindent\faSliders~\textbf{Disabling CoT.}
In our experiments, we evaluate models with explicit CoT disabled.
Following the practice adopted by \texttt{Qwen3-8B}~\cite{qwen3technicalreport}, we suppress intermediate reasoning by appending an empty thinking segment, i.e., ``\texttt{<think> </think>}'', to the input prompt.
This operation prevents the model from emitting explicit reasoning traces while preserving its ability to produce final answers.

\noindent\faSliders~\textbf{Context Length and Prompt Control.}
To ensure fair comparison, the maximum context length is fixed to \textbf{16k} tokens for all models across all experiments.
To avoid prompt-induced variation in reasoning length, we exclude any instructions that explicitly encourage or constrain reasoning behavior (e.g., ``Please reason step by step'').

For example, for the HumanEval dataset, we use the following instruction:

\begin{leftlinebox}
    \textit{Please write a Python program to address the following QUESTION. Your ANSWER should be in a code block format like this: \texttt{\textasciigrave\textasciigrave\textasciigrave}python \# Write your code here \texttt{\textasciigrave\textasciigrave\textasciigrave}.}
\end{leftlinebox}

\noindent\faSliders~\textbf{Loop Detection.}
Following \citet{pipis2025reasoningloop}, we identify looping behavior using an $n$-gram repetition criterion.
A response is classified as looping if any $n$-gram appears at least $k$ times within the same output.
Unless otherwise specified, we set $n = 30$ and $k = 20$ for all reasoning models, consistent with the original study.
This criterion captures severe degenerative repetition patterns commonly observed under greedy or low-temperature decoding and is empirically robust to moderate variations of $n$ and $k$.

\subsection{Empirical Findings}

\noindent\faSearch~\textbf{How does CoT perform?}
Table~\ref{tab:empirical_results} shows the performance of CoT in different models. The results reveal that excessive CoT length is a common phenomenon across most evaluated models.
\dx{Except for \texttt{gpt-oss-20b}, most models generate long reasoning traces. Their average CoT length ranges from about 2K to over 10K tokens, depending on the benchmark.}
\dx{We also recorded significant truncation rates on HumanEval, MBPP-Sanitized, and particularly LiveCodeBench, where the rate on \texttt{DeepSeek-Qwen-7B} reached 52.7\%.}
The results suggest that excessive CoT verbosity is a common issue in modern reasoning models. It appears consistently across diverse architectures and model sizes, frequently exceeding generation budgets.

  \begin{table}[htbp]
  \centering
  \caption{Performance Comparison on HumanEval, MBPP-Sanitized and LiveCodeBench (Max Generated Tokens = 16,384)}
  \label{tab:empirical_results}
  \begin{threeparttable}
  \begin{tabular}{@{}c@{}}
  \resizebox{\textwidth}{!}{
  \rowcolors{6}{gray!15}{white}
  \begin{tabular}{lccccccccc}
  \toprule
  \multirow{2}{*}{Models} & \multicolumn{3}{c}{HumanEval (164 Questions)} & \multicolumn{3}{c}{MBPP-Sanitized (378 Questions)} & \multicolumn{3}{c}{\dx{LiveCodeBench (1055 Questions)}} \\
  \cmidrule(lr){2-4} \cmidrule(lr){5-7} \cmidrule(lr){8-10}
   & $pass@1(\%)$ & CoT / Total & Truncation \# & $pass@1(\%)$ & CoT / Total & Truncation \# & \dx{$pass@1(\%)$} & \dx{CoT / Total} & \dx{Truncation \#} \\
  \midrule
  Qwen3-8B (w/o CoT) & 84.76\% & - / 249 & 0 (0.0\%) & 82.01\% & - / 78 & 0 (0.0\%) & \dx{40.57\%} & \dx{- / 2,162} & \dx{123 (11.7\%)} \\
  Qwen3-8B & 89.02\% & 3,604 / 3,701 & 11 (6.7\%) & 84.13\% & 3,663 / 3,712 & 43 (11.4\%) & \dx{61.61\%} & \dx{9,185 / 9,384} & \dx{329 (31.2\%)} \\
  \midrule
  DeepSeek-Qwen-7B (w/o CoT) & 65.24\% & - / 412 & 0 (0.0\%) & 50.79\% & - / 173 & 0 (0.0\%) & \dx{25.78\%} & \dx{- / 860} & \dx{11 (1.0\%)} \\
  DeepSeek-Qwen-7B & 78.04\% & 4,099 / 4,487 & 28 (17.1\%) & 73.54\% & 2,459 / 2,806 & 33 (8.7\%) & \dx{32.50\%} & \dx{10,229 / 10,500} & \dx{556 (52.7\%)} \\
  \midrule
  DeepSeek-Qwen-14B (w/o CoT) & 82.32\% & - / 456 & 0 (0.0\%) & 74.87\% & - / 424 & 0 (0.0\%) & \dx{36.87\%} & \dx{- / 1,245} & \dx{37 (3.5\%)} \\
  DeepSeek-Qwen-14B & 87.20\% & 3,282 / 3,669 & 18 (11.0\%) & 86.51\% & 2,078 / 2,453 & 17 (4.5\%) & \dx{49.95\%} & \dx{8,833 / 9,224} & \dx{366 (34.7\%)} \\
  \midrule
  DeepSeek-Qwen-32B (w/o CoT) & 87.20\% & - / 183 & 0 (0.0\%) & 88.62\% & - / 199 & 0 (0.0\%) & \dx{43.12\%} & \dx{- / 1,124} & \dx{14 (1.4\%)} \\
  DeepSeek-Qwen-32B & 88.41\% & 3,136 / 3,539 & 17 (10.4\%) & 88.10\% & 2,333 / 2,686 & 28 (7.4\%) & \dx{48.44\%} & \dx{9,137 / 9,478} & \dx{438 (41.5\%)} \\
  \midrule
  \dx{Qwen3.5-122B (w/o CoT)} & \dx{96.34\%} & \dx{- / 247} & \dx{0 (0.0\%)} & \dx{91.80\%} & \dx{- / 292} & \dx{2 (0.5\%)} & \dx{73.08\%} & \dx{- / 2,448} & \dx{70 (6.6\%)} \\
  \dx{Qwen3.5-122B} & \dx{96.95\%} & \dx{2,561 / 2,750} & \dx{4 (2.4\%)} & \dx{93.12\%} & \dx{1,479 / 1,620} & \dx{4 (1.1\%)} & \dx{77.25\%} & \dx{9,548 / 9,882} & \dx{245 (23.2\%)} \\
  \midrule
  \dx{Qwen3.5-397B (w/o CoT)} & \dx{96.34\%} & \dx{- / 297} & \dx{0 (0.0\%)} & \dx{93.39\%} & \dx{- / 264} & \dx{0 (0.0\%)} & \dx{77.82\%} & \dx{- / 1,994} & \dx{19 (1.8\%)} \\
  \dx{Qwen3.5-397B} & \dx{95.73\%} & \dx{6,052 / 6,257} & \dx{19 (11.6\%)} & \dx{95.24\%} & \dx{5,409 / 5,531} & \dx{25 (6.6\%)} & \dx{83.70\%} & \dx{9,691 / 10,070} & \dx{200 (19.0\%)} \\
  \midrule
  \midrule
  \rowcolor{white}
  DeepSeek-Llama-8B & 76.22\% & 3,640 / 4,021 & 19 (11.6\%) & 63.23\% & 2,568 / 2,910 & 35 (9.3\%) & \dx{33.27\%} & \dx{10,533 / 10,808} & \dx{542 (51.4\%)} \\
  QwQ-32B & 93.90\% & 2,363 / 2,458 & 2 (1.2\%) & 93.65\% & 1,929 / 1,990 & 6 (1.6\%) & \dx{70.81\%} & \dx{7,982 / 8,153} & \dx{216 (20.5\%)} \\
  gpt-oss-20b & 92.68\% & 481 / 790 & 1 (0.6\%) & 94.44\% & 470 / 693 & 3 (0.8\%) & \dx{74.03\%} & \dx{2,964 / 3,298} & \dx{93 (8.8\%)} \\
  \bottomrule
  \end{tabular}}
  \end{tabular}
  \begin{tablenotes}[para,flushleft]
      \item[*] \texttt{DeepSeek-Llama-8B}, \texttt{QwQ-32B} and \texttt{gpt-oss-20b} do not support disabling CoT.
  \end{tablenotes}
  \end{threeparttable}
  \end{table}

To evaluate the impact of CoT, we compare the results with and without CoT on models that support disabling CoT reasoning.
\dx{Enabling CoT improves pass@1 by 6.6\% on average across the evaluated settings, showing that CoT can improve reasoning performance.}
Specifically, the performance gains are substantially larger for smaller models (e.g., +12.8\% on HumanEval for \texttt{DeepSeek-Qwen-7B}), indicating that CoT plays a more critical role when model capacity is limited.
\dx{On simpler benchmarks such as HumanEval, the performance difference between models with and without CoT is small for large models with 32B or more parameters. For example, \texttt{DeepSeek-Qwen-32B} improves by only 1.21\%, and \texttt{Qwen3.5-122B} by 0.61\%. However, on the more complex LiveCodeBench benchmark, CoT still provides significant gains, such as +21.04\% for \texttt{Qwen3-8B}, and +5.88\% for \texttt{Qwen3.5-397B}.}

However, these accuracy improvements come at a substantial efficiency cost.
Enabling CoT increases the token cost by up to an order of magnitude in some cases (e.g., from approximately 400 tokens to over 4K tokens for \texttt{DeepSeek-Qwen-7B}), \dx{which suggests much higher inference overhead}.
\dx{This token overhead leads to significant latency. For \texttt{DeepSeek-Qwen-32B} on LiveCodeBench, the average per-problem inference time increases from 40s without CoT to 339s with CoT, based on a measured generation speed of 28 tokens per second on an A100 GPU. Such latency is impractical for most users.}
Moreover, the long CoT frequently triggers truncation, which reduces the accuracy improvements from CoT.
\dx{This issue is more evident on simpler tasks, where the limited gains from CoT can be offset by looping and truncation.}
\dx{In some cases, CoT even hurts accuracy: for example, \texttt{DeepSeek-Qwen-32B} drops from 88.62\% to 88.10\% on MBPP-Sanitized, and \texttt{Qwen3.5-397B} drops from 96.34\% to 95.73\% on HumanEval.}
\dx{By contrast, \texttt{gpt-oss-20b} shows the value of short CoT: it achieves 92.68\% pass@1 on HumanEval with only 481 CoT tokens on average. It outperforms the dense \texttt{DeepSeek-Qwen-32B} model and approaches the much larger \texttt{Qwen3.5-122B}. This result suggests that effective reasoning does not necessarily require long CoT. Instead, concise reasoning can preserve strong performance while reducing token cost and truncation risk.}

Overall, these results highlight a fundamental trade-off in the use of CoT.
While CoT improves reasoning accuracy, its effectiveness is constrained by excessive verbosity and failures caused by truncations. 
Taken together, these findings suggest that the key challenge is how to retain the benefits of CoT while avoiding unnecessary reasoning overhead.

\begin{center}
    \resizebox{\linewidth}{!}{
\begin{tabular}{l!{\vrule width 1pt}p{0.9\columnwidth}}
    \makecell{{\LARGE \faLightbulbO}}  & \textbf{\newfinding}
    \textbf{CoT improves accuracy but \dx{also increases token cost and truncation risk}.}
    \dx{Across models and benchmarks, enabling CoT improves pass@1 accuracy by 6.6\% on average, with larger gains on more complex tasks such as LiveCodeBench. However, it can increase token cost by about 10 times and often causes truncation, with the highest rate reaching 52.7\%.}
\end{tabular}}
\end{center}

\noindent\faHandORight~\textbf{How Common is Looping among Truncations?}
To quantify how often truncation is associated with looping behavior, we analyze all truncated outputs using the loop detection criterion described in Section~\ref{sec:empirical_implement_detail}.
Across all \dx{3,298} truncation cases with CoT observed across different models and datasets, \textbf{2,848 cases (86.4\%)} are identified as looping by the $n$-gram repetition criterion.

Since our loop detector is intentionally strict ($n{=}30$, $k{=}20$), borderline cases with weaker or shorter repetition may be missed.
We manually examine the remaining truncation cases not flagged by the detector and find that some of them still exhibit loop-like repetition patterns, although they do not satisfy the strict $n$-gram threshold.
Therefore, the detected 86.4\% should be viewed as a conservative lower bound, and the true fraction of truncations caused by looping is likely higher.

\begin{center}
    \resizebox{\linewidth}{!}{
\begin{tabular}{l!{\vrule width 1pt}p{0.9\columnwidth}}
    \makecell{{\LARGE \faLightbulbO}}  & \textbf{\newfinding\label{finding:truncation}}
    \textbf{Looping is highly common in truncated outputs.}
    Using a strict $n$-gram repetition detector, 86.4\% of truncations are identified as looping, and manual inspection suggests this is a conservative estimate.
\end{tabular}}
\end{center}

\noindent\faSearch~\textbf{How Much Reasoning is Effective?}
\dx{To estimate the useful part of a CoT,} we conduct an LLM-based CoT compression analysis. Following prior work that employs strong LLMs as judges or analyzers to evaluate model-generated text~\cite{NEURIPS2023_91f18a12}, we use \texttt{GPT-4o}~\cite{gpt4o} to \dx{compress CoT outputs while keeping their main reasoning steps. Similarly, C3oT~\cite{kang_c3ot_2025} uses GPT-4 to shorten CoTs while keeping key information and interpretability.}
Specifically, this process  \dx{aims to remove repeated statements, unnecessary steps, and low-value explanatory details while keeping the main reasoning steps and planning instructions.} \dx{We use this analysis only as an approximate estimate. It does not show which tokens are strictly needed, nor does it imply that the compressed trace behaves the same as the full CoT.}
The prompt used for this section is provided in the replication package.
To reduce randomness in LLM-based compression, we perform the compression three times for each CoT and report the averaged results.

After compression, we measure 1) the token length of the compressed CoT and 2) the effective reasoning ratio, defined as the fraction of tokens retained after compression relative to the original CoT length. The results are reported in Table \ref{tab:cot_compression_gpt4o}. 
Truncated instances are omitted because they have not finished the reasoning process.

\begin{table}[htb]
\centering
\caption{Effective Reasoning Length and Retention Ratio Estimated via GPT-4o on HumanEval and MBPP}
\label{tab:cot_compression_gpt4o}
\rowcolors{5}{white}{gray!15}
\resizebox{\textwidth}{!}{
\begin{tabular}{lccc ccc}
\toprule
\multirow{2}{*}{Models} & \multicolumn{3}{c}{HumanEval (164 Questions)} & \multicolumn{3}{c}{MBPP-Sanitized (378 Questions)} \\
\cmidrule(lr){2-4}\cmidrule(lr){5-7}
& Avg. CoT (Before) 
& Avg. CoT (After) 
& Retained (\%) 
& Avg. CoT (Before) 
& Avg. CoT (After) 
& Retained (\%) \\
\midrule
DeepSeek-Llama-8B & 3,594 & 212 & 5.91  & 2,538 & 164 & 6.50  \\
DeepSeek-Qwen-7B  & 3,677 & 210 & 5.71  & 2,334 & 177 & 7.62  \\
DeepSeek-Qwen-14B & 3,076 & 211 & 6.88  & 1,974 & 181 & 9.17  \\
DeepSeek-Qwen-32B & 2,869 & 212 & 7.42  & 2,218 & 179 & 8.07  \\
QwQ-32B           & 2,311 & 216 & 9.36  & 1,866 & 187 & 10.05 \\
Qwen3-8B          & 3,456 & 228 & 6.62  & 3,527 & 155 & 4.41  \\
gpt-oss-20b       &   481 & 149 & 31.11 &   470 & 128 & 27.27 \\
\bottomrule
\end{tabular}
}
\end{table}

Table~\ref{tab:cot_compression_gpt4o} \dx{suggests that generated CoTs contain much redundant content in this compression analysis}. For most models (except \texttt{gpt-oss-20b}), the compressed lengths consistently fall between 150 and 230 tokens across benchmarks. Consequently, the retained-token ratio falls below 10\%, suggesting that \dx{a large fraction of generated tokens may not be needed for the core reasoning under this analysis}. Additionally, \texttt{gpt-oss-20b} \dx{appears more concise, although the same compression process still removes roughly 70\% of its tokens}.

The comparison results between \texttt{Qwen3-8B} and \texttt{QwQ-32B} on MBPP-Sanitized further illustrates this phenomenon. Although \texttt{Qwen3-8B} generates nearly twice as many CoT tokens as \texttt{QwQ-32B} before compression, its \dx{retained-token} ratio after compression is even lower. Under this analysis, increased CoT verbosity does not correspond to more effective reasoning and may instead amplify redundant or low-value content. \dx{Overall, these results suggest that only a small portion of CoT contains core effective information, making CoT compression necessary for producing concise reasoning traces.}

\begin{center}
    \resizebox{\linewidth}{!}{
\begin{tabular}{l!{\vrule width 1pt}p{0.9\columnwidth}}
    \makecell{{\LARGE \faLightbulbO}}  & \textbf{\newfinding}
    \textbf{\dx{Using an LLM-based compression analysis, only a small fraction of CoT tokens is retained.}}
    Compression-based analysis with GPT-4o shows that, despite large variation in raw CoT length, less than 10\% of CoT tokens are retained across most models. \dx{We treat this as evidence of large redundancy, but not as an exact measure of which reasoning tokens are truly needed.}
\end{tabular}}
\end{center}

\begin{figure*}[t] 
  \centering
  \includegraphics[width=0.9\textwidth]{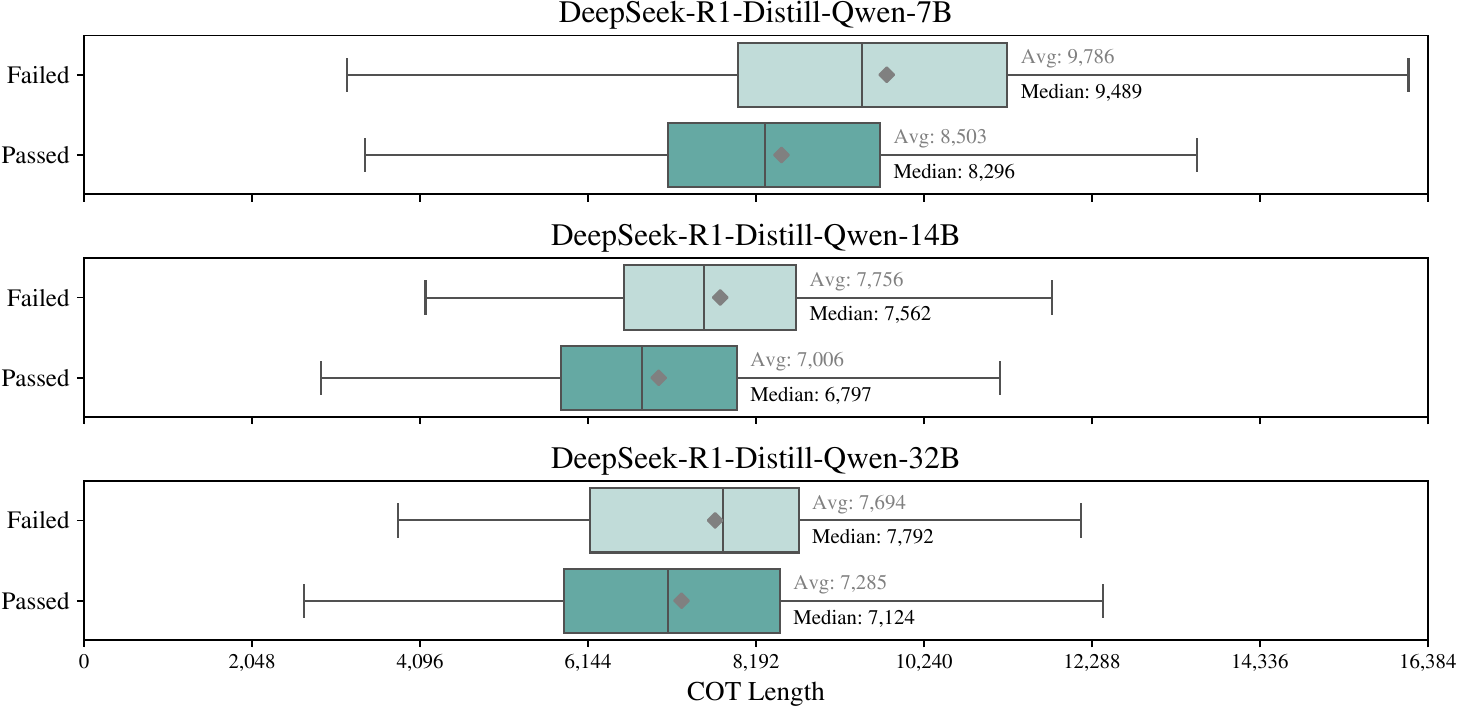} 
  \caption{Distribution of CoT lengths for passed and failed cases across DeepSeek-R1-Distill-Qwen models (7B, 14B, and 32B) on \textbf{HumanEval/129}. This case study suggests that successful outputs do not necessarily require longer CoT reasoning, and that excessively long reasoning chains can \dx{be linked to worse outcomes in this case}.} 
  \label{fig:cot} 
\end{figure*}

\noindent\faSearch~\textbf{Is Longer Reasoning Always Better?}
To evaluate whether longer CoT reasoning leads to better outcomes, we generate multiple responses for a single problem and analyze the CoT lengths of passed and failed solutions.  Since HumanEval does not provide an official difficulty classification, we select the problem with the longest prompt in the dataset, i.e., \textbf{HumanEval/129}. \dx{We note that this is a single-case analysis. It provides qualitative evidence rather than a benchmark-level conclusion.}
We choose the \texttt{DeepSeek-Qwen-Series} Model for our evaluation, focusing on how the length of CoT impacts its performance across various model sizes.
The prompt follows the implementation details of this section (Section~\ref{sec:empirical_implement_detail}), without any explicit instructions to control the length of the CoT.
For each of the models, we generate 10,000 responses and examine the distribution of CoT lengths, as shown in Figure~\ref{fig:cot}, with the maximum token length set to 16,384.

While prior work~\cite{jin2024impact} suggests that increasing the length of CoT reasoning improves accuracy, this case study reveals a different pattern across all three evaluated models.
Failed outputs are frequently associated with longer CoT lengths, while successful outputs are more concentrated around shorter lengths.
For instance, in the \texttt{7B} model, the median CoT length of failed generations (9{,}489 tokens) surpasses that of successful generations (8{,}296 tokens). Similar pattern observed in both \texttt{14B} and \texttt{32B} models. 
\dx{We further test another randomly selected problem, \textbf{HumanEval/14}, which is much easier, as it has a substantially higher pass@1 in our runs. The same pattern appears more clearly: for \texttt{DeepSeek-Qwen-7B}, \texttt{14B}, and \texttt{32B}, failed runs are much longer than successful runs on average (2{,}037 vs.\ 850, 2{,}406 vs.\ 793, and 1{,}851 vs.\ 755 tokens, respectively). This additional case suggests that overlong CoT is often redundant rather than necessary for solving the task.}
A closer examination of failure cases \dx{suggests that overly long reasoning can add redundancy, propagate errors, and cause truncation, which may hurt performance}.
\dx{Overall, this case study suggests} that effective reasoning does not simply require ``more thinking'', but rather \dx{an appropriate reasoning length} that balances sufficiency and conciseness.

\begin{center}
    \resizebox{\linewidth}{!}{
\begin{tabular}{l!{\vrule width 1pt}p{0.9\columnwidth}}
    \makecell{{\LARGE \faLightbulbO}}  & \textbf{\newfinding\label{finding4}}
    \textbf{\dx{A single-case study suggests that longer CoT can be linked to failures.}}
    On HumanEval/129, failed cases show longer CoT than successful ones across all \dx{evaluated models} (e.g., 9{,}489 vs.\ 8{,}296 tokens in 7B), \dx{suggesting that excessive reasoning can be linked to failures in this case}.
\end{tabular}}
\end{center}

\noindent\faSearch~\textbf{Can Prompt Design Effectively Reduce CoT Length?}
Prior work has shown that the verbosity of CoT reasoning is sensitive to instruction-level guidance~\cite{wei2022chain,wang2023selfconsistency,agrawal2024taming}. Therefore, to verify whether the prompt design can effectively shorten the length of the CoT, we include a lightweight prompting baseline to encourage concise reasoning.
Specifically, we augment the original task instruction with a general conciseness guideline, asking the model to provide its reasoning as concisely as possible while avoiding redundant or unnecessary steps.
Concretely, we add the following system prompt:

\begin{leftlinebox}
    \textit{Please provide your reasoning as concisely as possible, avoiding redundant or unnecessary steps.}
\end{leftlinebox}

\begin{table}[htbp]
\centering
\caption{Effect of Conciseness-Guided Prompting on CoT Length and Model Accuracy}
\label{tab:cot_length_accuracy}
\rowcolors{5}{white}{gray!15}
\resizebox{\textwidth}{!}{
\begin{tabular}{lcccccc}
\toprule
\multirow{2}{*}{Models} & \multicolumn{3}{c}{HumanEval (164 Questions)} & \multicolumn{3}{c}{MBPP-Sanitized (378 Questions)} \\
\cmidrule(lr){2-4} \cmidrule(lr){5-7}
 & Original CoT & Concise CoT ($\Delta$Len) & Acc. $\Delta$ & Original CoT & Concise CoT ($\Delta$Len) & Acc. $\Delta$ \\
\midrule
DeepSeek-Llama-8B & 3,640 & 3,063 (-15.9\%) & +2.44\% & 2,568 & 2,564 (-0.2\%)  & -1.85\% \\
DeepSeek-Qwen-7B  & 3,839 & 3,464 (-9.8\%)  & +0.61\% & 2,459 & 1,946 (-20.9\%) & -5.55\% \\
DeepSeek-Qwen-14B & 3,282 & 2,890 (-11.9\%) & +1.82\% & 2,078 & 1,952 (-6.1\%)  & -1.06\% \\
DeepSeek-Qwen-32B & 3,136 & 2,610 (-16.8\%) & +1.83\% & 2,333 & 1,890 (-19.0\%) & +1.58\% \\
QwQ-32B           & 2,363 & 2,236 (-5.4\%)  & 0.00\%  & 1,929 & 1,779 (-7.8\%)  & +0.53\% \\
Qwen3-8B          & 3,604 & 2,490 (-30.9\%) & +3.05\% & 3,663 & 2,545 (-30.5\%) & +3.70\% \\
gpt-oss-20b       & 481   & 238 (-50.5\%)   & +4.88\% & 470   & 241 (-48.7\%)   & -2.64\% \\
\bottomrule
\end{tabular}
}
\end{table}

As shown in Table~\ref{tab:cot_length_accuracy}, introducing this prompt instruction reduces the average CoT length for most evaluated models. This confirms that CoT verbosity is sensitive to instruction-level guidance. However, the reduction rate varies substantially across models and datasets, indicating that prompt-based control lacks reliability and offers limited effectiveness in practice.

In particular, while moderate reductions are observed for several DeepSeek and Qwen models, the overall compression ratios are relatively small and highly model-dependent. 
In contrast, \texttt{gpt-oss-20b} exhibits a much more pronounced reduction in reasoning length (over 50\% on HumanEval). 
One possible explanation is that \texttt{gpt-oss-20b} natively supports explicit control of reasoning length, whereas most other open-source models lack such mechanisms, making them less responsive to high-level conciseness instructions.

It should be noted that shortening CoT via prompting does not uniformly degrade performance. In several cases, accuracy remains stable or even improves after applying the concise-CoT prompt (e.g., \texttt{Qwen3-8B} and \texttt{DeepSeek-Qwen-32B} on HumanEval). This observation \dx{is consistent with our HumanEval/129 case study}, suggesting that excessively long CoT does not necessarily lead to better reasoning. On the contrary, overly verbose reasoning may introduce redundant or noisy content that hinders effective problem-solving.
Overall, prompt design can influence CoT length but lacks robustness and consistency across models. The fact that shorter CoT can maintain or even improve accuracy shows that reducing unnecessary reasoning is both feasible and useful.

\begin{center}
\resizebox{\linewidth}{!}{
\begin{tabular}{l!{\vrule width 1pt}p{0.9\columnwidth}}
\makecell{{\LARGE \faLightbulbO}}  &
\textbf{\newfinding\label{finding:prompt}}
\textbf{Prompt-based conciseness control is limited and model-dependent.}
Conciseness prompts reduce CoT length for most models, but the gains are generally small and inconsistent.
\dx{Even with a conciseness prompt, CoT length is still hard to control across models and datasets.}
\end{tabular}}
\end{center}

\section{Methodology}\label{sec:method}
\begin{figure*}[t] 
  \centering
  \includegraphics[width=0.98\textwidth]{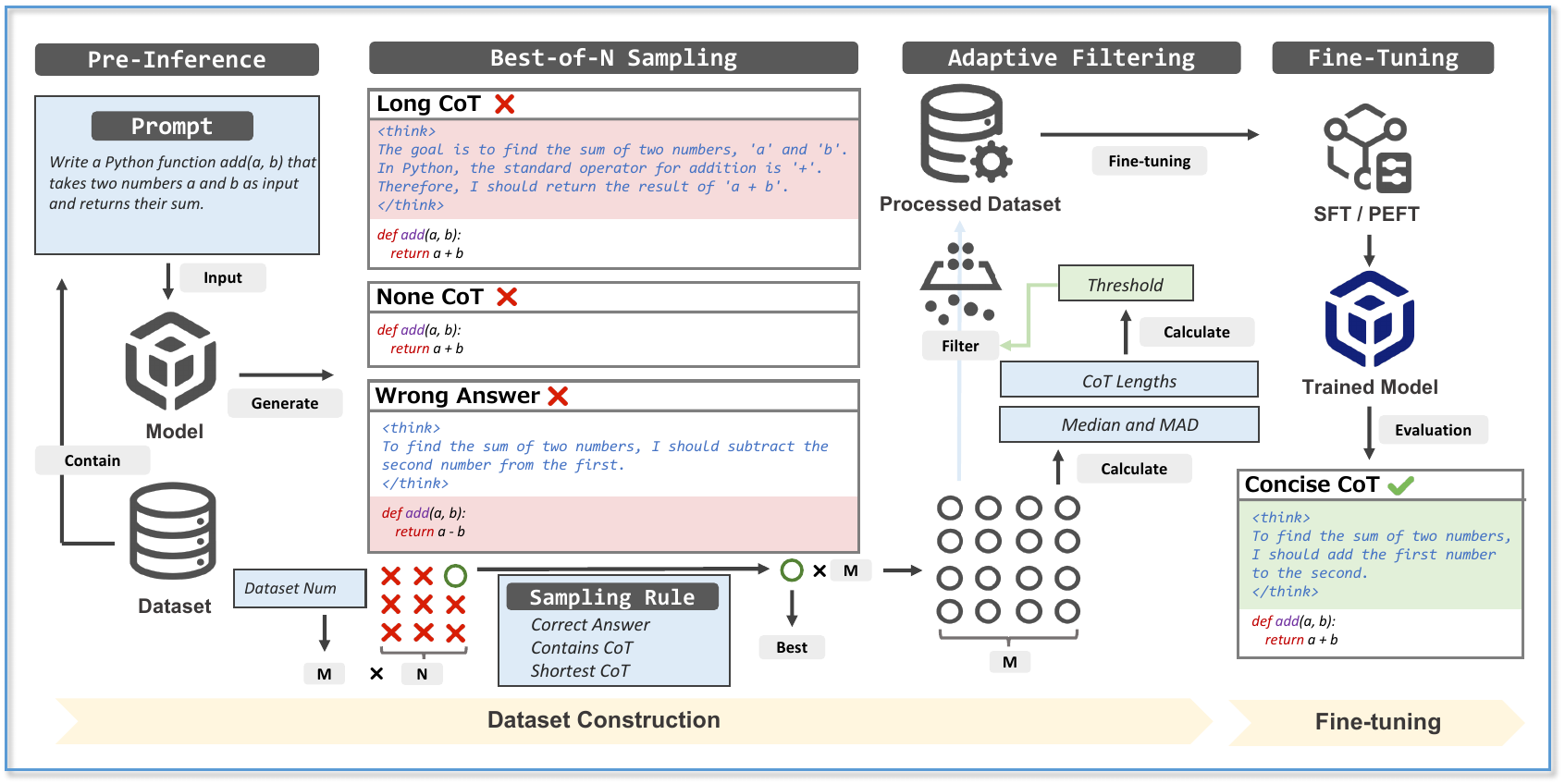} 
  \caption{An overview of the \name framework, illustrating its self-enhancing procedure of generation, selection, and adaptive filtering.} 
  \label{fig:structure} 
\end{figure*}

We propose \name, a self-enhancing framework designed to enable LLMs to refine their reasoning capabilities by learning from their own generated outputs. 
Unlike conventional supervised fine-tuning pipelines that depend heavily on high-quality, human-labeled CoT annotations, \name operates in a fully autonomous manner: the model generates candidate reasoning traces, filters them based on correctness and conciseness, and then learns from the selected high-quality examples. As illustrated in Figure~\ref{fig:structure}, the framework is composed of three key stages: 
(1) \textbf{Pre-inference generation} of CoT responses, 
(2) \textbf{BoN sampling} for high-quality reasoning path selection, and 
(3) \textbf{Adaptive CoT filtering} to control verbosity without sacrificing reasoning fidelity.
Through this design, \name significantly reduces the reliance on prompt engineering and external compression tools, while promoting reasoning traces that are both accurate and concise, and that remain tailored to the nature of each target task.

\subsection{Pre-Inference Data Generation}
The first stage of \name focuses on constructing a comprehensive corpus of CoT outputs to serve as the foundation for subsequent filtering and fine-tuning. In this step, the base language model is used to process each question in the training set, generating a complete solution that includes an explicit, step-by-step reasoning path.
This step does not rely on prompt design. Therefore, we intentionally avoid using complex prompts. For example, for code generation tasks, we adopt the same prompt as in Section~\ref{sec:empirical_implement_detail}.
To avoid early truncation that may prevent the model from completing its reasoning and based on the reasoning length observed in our empirical study, we adopt a moderate token budget during generation, i.e., 16k. Moreover, this setting can help collect more complete and diverse reasoning traces, thereby reducing information loss and mitigating overfitting caused by insufficient CoT data.

\subsubsection{Best-of-N Sampling for Data Refinement}
Since reasoning loops are a common phenomenon in CoT generation for reasoning models, we adopt a \textbf{BoN} sampling strategy to mitigate looping behaviors. In addition, this strategy enables us to collect more diverse reasoning traces for constructing a high-quality training set, and can be used to expand the dataset when the available training data are limited.

Specifically, for each input question, the model generates N candidate responses, each accompanied by a CoT. These candidates are then ranked and filtered according to a set of criteria:
\ding{182} \textbf{Correctness of the Final Answer:} only candidates producing the correct final answer are considered. All incorrect outputs are discarded.
\ding{183} \textbf{Presence of Valid CoT:} from the remaining candidates, we retain only those containing a non-empty and non-looping reasoning path. \dx{This helps each retained example remain interpretable and suitable for supervised training.}
\ding{184} \textbf{Conciseness of Reasoning:} if multiple candidates satisfy the first two criteria, we select the one with the shortest CoT length. This prioritizes efficiency and prevents the inclusion of unnecessarily verbose reasoning.

\subsection{Adaptive CoT Filtering}
After BoN sampling, CoT lengths remain highly variable and often contain long-tail verbosity.
To obtain concise yet correct training signals, we apply a \textbf{two-stage filter}.

\noindent\faFilter~\textbf{Stage 1: Output validity.}
We first remove samples whose final outputs are invalid (e.g., code that cannot be parsed or fails the provided unit tests in code-generation datasets).
This step filters out malformed or incomplete outputs before applying any length-based filtering.

\noindent\faFilter~\textbf{Stage 2: Robust length filter.}
On the remaining samples, we filter overly long CoTs using a robust, distribution-aware threshold.
This is motivated by the empirical observation that CoT lengths are often heavy-tailed: a small fraction of samples can be extremely long due to overthinking or looping, and these outliers can dominate both training cost and the learned behavior.

Let $\lambda_i$ denote CoT length, and let $\tilde{\lambda}=\mathrm{median}(\{\lambda_i\})$ be the median length.
We compute the median absolute deviation (MAD), $\mathrm{MAD}=\mathrm{median}(|\lambda_i-\tilde{\lambda}|)$, and define the cutoff as
$\lambda_{\mathrm{mad}}(\alpha)=\tilde{\lambda}+\alpha\cdot\mathrm{MAD}$,
where $\alpha$ controls the strictness of filtering.
By default, we use $\alpha=1$ (i.e., $\lambda_c=\lambda_{\mathrm{mad}}(1)$) and discard samples with $\lambda_i>\lambda_c$.
Compared with mean-based thresholds, this rule is less sensitive to extreme outliers and yields a stable cutoff.

\section{Experimental Setup}\label{sec:setup}
\subsection{Datasets}
To evaluate the performance and effectiveness of our method, we conduct experiments on a set of representative software engineering datasets. The experiments cover diverse task formats and reasoning requirements. 

\subsubsection{Training and Evaluation Benchmarks.}
We use \textbf{MathQA-Python}~\cite{austin2021program}, \textbf{CodeXGLUE-Defect-Detection}~\cite{zhou2019devign}, \textbf{Code-Search-WebQuery}~\cite{Lu2021CodeXGLUEAM,husain2019codesearchnet}, \dx{and \textbf{APPS}~\cite{hendrycksapps2021}}. All of these datasets provide predefined training and test splits.
\textbf{MathQA-Python} focuses on generating executable Python programs to solve mathematically formulated problems, requiring both symbolic reasoning and correct code implementation.
\textbf{CodeXGLUE-Defect-Detection} is a binary classification task that assesses whether a given code snippet contains functional defects or security vulnerabilities, emphasizing program understanding and static reasoning.
\textbf{Code-Search-WebQuery} evaluates semantic code retrieval by determining whether a code snippet satisfies a natural language query, challenging the model to align program semantics with textual descriptions.
\dx{\textbf{APPS} is a code-generation benchmark built from natural-language programming problems and test cases. It requires the model to produce executable solutions in more diverse coding settings.}
These four datasets are used to train \dx{\name and all trainable baselines with the same data splits and optimization settings, while prompting-only baselines are evaluated without additional training}. Evaluation is conducted on  the respective test sets.

\subsubsection{Evaluate-Only Benchmarks.}
In addition, we evaluate our method on \textbf{HumanEval} and \textbf{MBPP-Sanitized}, which are used exclusively for evaluation.
These datasets are not involved in training, and serve to assess the generalization capability of our method on unseen code generation tasks.

Overall, this benchmark suite enables a systematic evaluation of both in-domain performance and cross-task generalization across software engineering scenarios.

\subsection{Implementation Details}
All experiments are conducted on a server equipped with four NVIDIA RTX A800 GPUs (80 GB each) and 128 Intel Xeon CPUs, running Ubuntu 22.04 LTS. 
We fine-tune \texttt{DeepSeek-R1-Distill-Qwen-7B} model~\cite{deepseekai2025deepseekr1} under a \textbf{full-parameter supervised fine-tuning (SFT)} setting, using the LLaMAFactory framework~\cite{zheng2024llamafactory}. \dx{Unless otherwise stated, we use the same backbone for pre-inference candidate generation, fine-tuning, and final evaluation.}
Training is performed for \textbf{three epochs} with an effective batch size of eight, gradient checkpoint enabled, and the \textbf{AdamW} optimizer is used. 
The learning rate is fixed at $1 \times 10^{-5}$ with a cosine decay schedule and 10\% warm-up, and gradient clipping (norm 1.0) is applied for stability. 
During fine-tuning, inputs are tokenized with a maximum sequence length of 16{,}384 tokens. 
Although we adopt full-parameter SFT by default, the \name framework is also compatible with parameter-efficient fine-tuning methods such as LoRA and QLoRA, which we discuss in the Discussion section. 

\subsection{Baselines} 
In our work, we consider the following baselines.

\begin{itemize}[leftmargin=*]
    \item \textit{Short CoT}: Employ prompting techniques to elicit a more concise CoT from the model, using the prompt design described in Section~\ref{sec:empirical}.
    \item \textit{Plain SFT}: \dx{A fine-tuning baseline that uses the same self-generated training data and training schedule as \name, but without BoN refinement or adaptive length filtering.}
    \item \textit{TokenSkip}~\cite{xia_tokenskip_2025}: Fine-tuning LLMs by selectively skipping less important tokens while retaining key tokens, enabling the generation of compressed CoTs with adjustable ratios.
    \item \textit{Naive BoN}~\cite{munkhbat2025selftraining}: Use only the Best-of-N method to process the pre-inference data and then train the model.
\end{itemize}

\subsection{Evaluation Metrics}
We measure the following metrics:

\textbf{Pass@1:} Pass@1 measures single-attempt correctness. Generated code is
correct if it passes all test cases, while classification and retrieval outputs
are correct if they match the ground-truth label.

\textbf{Compression Rate:} We define the compression rate as
$\rho = 1-\lambda'/\lambda$, where $\lambda$ and $\lambda'$ denote the token
lengths of the original and generated CoTs, respectively. A higher $\rho$
indicates stronger compression.

\section{Experiment}\label{sec:evaluation}
In this section, we conduct experiments to answer the following research questions.

\subsection{Research Questions} \label{RQ}
In order to evaluate our work, we address the following research questions:
\begin{itemize}[leftmargin=*]
    \item \textbf{RQ1: How does the CoT compression performance of \name compare with baselines?} We evaluate \name against baseline methods on multiple datasets.  
    \item \textbf{RQ2: How \dx{robust is \name across unseen benchmarks, backbones, and training approaches?}} \dx{We examine \name from three perspectives: evaluation on unseen benchmarks with the same backbone, results on other backbones, and behavior under different training approaches}.    
    \item \textbf{RQ3: What is the contribution of each component of \name?} We conduct ablation studies by removing or varying parts of \name, such as the length-filter strictness $\alpha$ (which determines $\lambda_c$) and the BoN parameter $\mathcal{N}$, to examine their impact.
    \item \textbf{RQ4: How can \name mitigate reasoning loops?} We test whether \name helps LLMs generate valid answers without encountering looping problems. 
\end{itemize}

\subsection{RQ1: The Performance of \name}

In this subsection, we evaluate the effectiveness of \name under the 16K token constraint by comparing it with baselines. 
For baselines with tunable parameters, we adopt commonly used and representative settings.
Specifically, TokenSkip is evaluated with $\gamma = 0.5$ (i.e., compression ratio) to examine its maximum compression behavior, and with $\gamma = 0.8$ to reflect a more conservative and widely used configuration.
For Naive BoN, we sample $N = 3$ candidates, which matches the sampling budget used in \name.
For BoN section in \name, we set the same parameter, i.e., $N = 3$.
\dx{We also include a matched \textit{Plain SFT} baseline that uses the same self-generated training data and optimization schedule as \name, but without BoN refinement or adaptive CoT filtering.}
The results are summarized in Table~\ref{tab:baselines}.

\begin{table}[t]
\begin{threeparttable}
\centering
\caption{The CoT Compression Performance Over Different Datasets (16K Token Limit)}
\label{tab:baselines}
\setlength{\tabcolsep}{4pt}
\begin{tabular}{@{}c@{}}
\resizebox{\linewidth}{!}{
\begin{tabular}{l ccc ccc ccc ccc c}
\toprule
\multirow{2}{*}{\textbf{Methods}} 
& \multicolumn{3}{c}{\textbf{MathQA-Python}} 
& \multicolumn{3}{c}{\textbf{Defect-Detection}} 
& \multicolumn{3}{c}{\textbf{Code-Search}} 
& \multicolumn{3}{c}{\dx{\textbf{APPS}}} 
& \multirow{2}{*}{\textbf{$\rho_{avg}(\%)$}} \\
\cmidrule(lr){2-4} \cmidrule(lr){5-7} \cmidrule(lr){8-10} \cmidrule(lr){11-13}
& $p@1(\%)$ & $\lambda_{avg}$ & $\rho(\%)$ 
& $p@1(\%)$ & $\lambda_{avg}$ & $\rho(\%)$ 
& $p@1(\%)$ & $\lambda_{avg}$ & $\rho(\%)$ 
& $p@1(\%)$ & $\lambda_{avg}$ & $\rho(\%)$
&  \\
\midrule[\heavyrulewidth]
Base                            
& 63.7 & 1,456 & - 
& 44.7 & 1,836 & - 
& 72.4 & 472   & - 
& \dx{30.8} & \dx{10,953} & \dx{-} 
& - \\
\midrule
TokenSkip ($\gamma$=0.5)        
& 1.6  & 1,196 & 17.9 
& 36.2 & 2,421 & -31.9 
& 66.5 & 465   & 1.5 
& \dx{0.0} & \dx{12,295} & \dx{-12.2} 
& \dx{-6.2} \\
TokenSkip ($\gamma$=0.8)        
& 0.9  & 1,109 & 23.8 
& 47.2 & 1,004 & 45.3 
& 76.4 & 362   & 23.3
& \dx{0.0} & \dx{11,914} & \dx{-8.8} 
& \dx{20.9} \\
\dx{Plain SFT}                    
& 74.6 & 1,177 & 19.2 
& 49.3 & 1,055 & 42.5 
& 72.5 & 468   & 0.9 
& \dx{36.6} & \dx{9,995} & \dx{8.7} 
& \dx{17.8} \\
Naive BoN (N=3)                 
& 73.6 & 1,191 & 18.2 
& 49.8 & 1,042 & 43.3 
& 74.9 & 478   & -1.3 
& \dx{37.1} & \dx{9,964} & \dx{9.0} 
& \dx{17.3} \\
Short CoT                       
& 55.6 & 1,236 & 15.1 
& 46.6 & 1,198 & 34.8 
& 69.7 & 398   & 15.7 
& \dx{31.7} & \dx{10,555} & \dx{3.6} 
& \dx{17.3} \\
\midrule
\textbf{\name}                  
& \textbf{74.9} & \textbf{877} & \textbf{39.8} 
& \textbf{50.5} & \textbf{785} & \textbf{57.2} 
& \textbf{77.3} & \textbf{341} & \textbf{27.8} 
& \dx{\textbf{37.5}} & \dx{\textbf{9,451}} & \dx{\textbf{13.7}} 
& \textbf{\dx{34.6}} \\
\bottomrule
\end{tabular}}
\end{tabular}
\begin{tablenotes}[para,flushleft]
    \item[*] The Base model refers to ``\texttt{DeepSeek-R1-Distill-Qwen-7B}''. Unless otherwise specified, this notation applies to all remaining tables in this paper; \\
    \item[*] $p@1(\%)$: Pass@1 accuracy; $\lambda_{avg}$: Average output length (tokens); \\
    \item[*] $\rho(\%)$: Compression rate compared to Base; $\rho_{avg}(\%)$: Average compression rate across MathQA-Python, Defect-Detection, Code-Search, and APPS; \\
    \item[*] $\gamma$ in TokenSkip: compression ratio, lower $\gamma$ means higher compression rate.
\end{tablenotes}
\end{threeparttable}
\end{table}

Overall, \name achieves the best performance in both reasoning efficiency and task accuracy across the four software engineering benchmarks.
On MathQA-Python, \name attains a pass@1 of 74.9\% while reducing the average CoT length by 39.8\%, indicating that substantial compression can be achieved without sacrificing correctness in code generation tasks.
On Defect-Detection, \name delivers the best overall performance, achieving the highest pass@1 (50.5\%) together with the most aggressive compression (57.2\%).
Similarly, on Code-Search, \name preserves the highest accuracy (77.3\%) while shortening reasoning traces by 27.8\%.
\dx{On APPS, \name also achieves the best result, reaching a pass@1 of 37.5\% and reducing the average output length from 10,953 to 9,451 tokens (13.7\% compression).}
These results demonstrate that \name can effectively control CoT verbosity while maintaining, or even improving, task performance across diverse settings.

In contrast, TokenSkip exhibits unstable behavior across compression settings. Under a low $\gamma$ setting (i.e., high compression rate), it performs particularly poorly: on Defect-Detection, the average CoT length even increases compared to the base model, leading to more truncation and lower accuracy. Although a higher $\gamma$ improves stability, TokenSkip still underperforms \name. Moreover, on MathQA-Python, its pass@1 drops to around 1\% regardless of the compression setting, as its token-level compression ignores code structure and produces unexecutable code.

Other trainable baselines without explicit compression operations, including  Plain SFT and Naive BoN, are more stable but yield limited improvements. \dx{Plain SFT is an important baseline. It improves the base model, yet it exhibits a lower compression rate and accuracy compared to \name on the evaluated datasets.} This suggests that \dx{the effectiveness of \name is attributed to fine-tuning, data refinement, and adaptive CoT filtering}.

Prompting-based methods also achieve a certain degree of CoT length reduction, with an average compression rate of 17.3\% across the four software engineering benchmarks. \dx{On APPS, Short CoT further reduces the average output length by 3.6\%, but the pass@1 gain is only 0.9 points (31.7\% vs.\ 30.8\%).} However, this approach comes at a noticeable cost to accuracy, as concise prompt may suppress necessary intermediate reasoning. On the coding task (i.e., MathQA-Python), pass@1 drops substantially by 8.1\%. These results further reinforce our Empirical Finding~\ref{finding:prompt}.

\dx{We also compare the resource cost of \name with the baselines. Since most baselines also run inference on the original training datasets, the main extra cost of \name comes from the additional inference for BoN sampling. With $N=3$, \name runs two extra inference passes. This cost is small compared with the following fine-tuning cost. Furthermore, by filtering out excessively long training samples, \name shortens the final training stage and reduces training time by approximately 30\% to 50\% in our experiments. When more training data are used, this filtering removes more redundant samples, making the resource benefit clearer. In addition, \name uses the same training and inference procedure as standard SFT, so it does not add extra CPU or memory requirements.}

\begin{center}
    \resizebox{\linewidth}{!}{
\begin{tabular}{l!{\vrule width 1pt}p{0.9\columnwidth}}
    \makecell{{\LARGE \faLightbulbO}}  &
    \textbf{Answer to RQ1:}
    \dx{Among the evaluated baselines, \name provides the best overall balance between accuracy and CoT compression across the four software engineering benchmarks.} On average, \name reduces CoT length by 34.6\%, exceeding the strongest baseline by 13.7\%, while maintaining or improving pass@1 accuracy under strict context constraints. \dx{On APPS, \name reaches the best pass@1 of 37.5\% with a 13.7\% reduction in average output length.}
\end{tabular}}
\end{center}

\subsection{RQ2: Generalization \dx{and Robustness} of \name}
In this section, we evaluate the robustness of \name \dx{from three perspectives: evaluation on unseen benchmarks, results on different backbones, and adaptation to different training approaches}. 

\subsubsection{\dx{Evaluation on Unseen Benchmarks}} 
\begin{figure*}[t] 
  \centering
  \includegraphics[width=\textwidth]{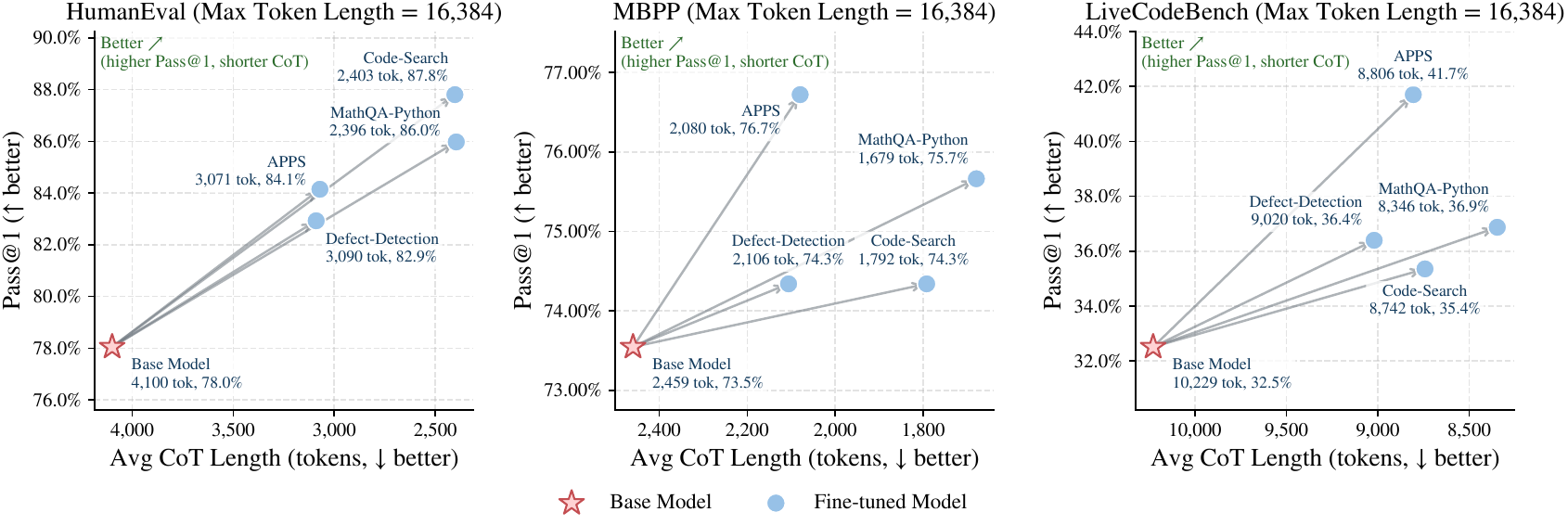}   
  \caption{Performance of \name fine-tuned on different software engineering datasets when evaluated on \textbf{HumanEval}, \textbf{MBPP} \dx{and \textbf{LiveCodeBench}}.
\dx{In this same-backbone evaluation setting, }\name  achieves higher accuracy while reducing the average length of CoT reasoning, compared to the base model.
}
  \Description{A scatter plot comparing accuracy and average chain-of-thought length on HumanEval, MBPP, and LiveCodeBench for the base model and three \name fine-tuned variants trained on different software engineering datasets. Across these benchmarks, the fine-tuned variants move toward higher accuracy and shorter reasoning length than the base model.}
  \label{fig:generalization} 
\end{figure*}

\dx{We use the four models fine-tuned on the four datasets in RQ1 and evaluate them on three unseen benchmarks, \textbf{HumanEval}, \textbf{MBPP}, and \textbf{LiveCodeBench}, as illustrated in Figure~\ref{fig:generalization}.}

\dx{Figure~\ref{fig:generalization} shows that, regardless of which dataset is used for fine-tuning, \name demonstrates strong cross-dataset generalization ability, improving both task accuracy and reasoning efficiency on unseen benchmarks.}
On \textbf{HumanEval}, the base model achieves about 78.0\% accuracy with an average CoT length of 4,100 tokens; the fine-tuned models improve accuracy by 4.9 to 9.8 points (e.g., from 78.0\% to 87.8\%) while reducing the average reasoning length to around 2,400 tokens (\,$\sim$40\% reduction).
On \textbf{MBPP}, the fine-tuned models improve accuracy by about 1.8 points (e.g., from 73.5\% to 76.7\%) and reduce the reasoning length to 1,679 to 2,100 tokens (up to 32\% reduction).
\dx{On \textbf{LiveCodeBench}, the fine-tuned models improve accuracy by 2.9 to 9.2 points (e.g., from 32.5\% to 41.7\%) and reduce the reasoning length to 8,346 to 9,020 tokens (up to 18\% reduction).}

Overall, \name achieves a stable CoT-length reduction on \dx{all three} benchmarks, suggesting that it learns generalizable compression behaviors within the evaluated backbone and task family.

\subsubsection{\dx{The Impact of Training Approaches}}\label{sec:training_approaches}
\dx{We further study whether \name works with different fine-tuning approaches. Beyond full-parameter supervised fine-tuning (SFT), we also test LoRA, a parameter-efficient method. The comparison shows two findings. First, SFT achieves the best overall performance, as expected. However, LoRA still retains most benefits of our method and improves both accuracy and compression over the base model. This result suggests that adaptive CoT filtering does not require updating all model parameters.}

\begin{table}[ht]
\centering
\caption{The Performance Comparison Between Using SFT and PEFT Fine-tuning Method}
\label{tab:peft}
\resizebox{0.85\columnwidth}{!}{
\begin{tabular}{lccc}
\toprule
           & \textbf{$pass@1(\%)$} & \textbf{Avg CoT Len $\lambda_{avg}$ (Tokens)} & \textbf{Compression Rate $\rho$ (\%)}   \\ \midrule
Base       & 63.67                                & 1,456                                          & \textbf{-}                           \\
\rowcolor{gray!15}
PEFT (LoRA) & 67.71                                & 949                                           & 34.8                                 \\ 
SFT        & \textbf{74.88}                       & \textbf{877}                                  & \textbf{39.8}                         \\
\bottomrule
\end{tabular}
}
\end{table}

\dx{Second, the LoRA-based approach shows that \name remains useful when resources are limited. LoRA reduces computation and storage costs, making it suitable when full SFT is too costly. Although the improvements are smaller than those of SFT, it still offers a good efficiency--effectiveness trade-off. Taken together, these results show that \name is effective under full SFT and remains flexible under limited-resource settings.}

\subsubsection{\dx{Performance on Other Backbones}}
\dx{We further test \name on three representative backbones covering different model families, architectures, and modalities: \textbf{GLM-Z1-9B-0414}, a GLM reasoning model with a scale close to our main backbone; \textbf{Qwen3-30B-A3B-Thinking-2507}, an MoE reasoning model for testing \name on an MoE structure; and \textbf{Qwen3.5-9B}, a recent multimodal dense model for testing whether \name remains effective on multimodal backbones.}

\dx{GLM-Z1-9B-0414 and Qwen3.5-9B are fine-tuned with full-parameter SFT, while Qwen3-30B-A3B-Thinking-2507 uses LoRA due to GPU-memory limits. As shown in Section~\ref{sec:training_approaches}, \name with LoRA performs worse than full-parameter SFT. Therefore, the MoE result reported here should be viewed as a conservative estimate.}

\begin{table}[ht]
\begin{threeparttable}
\centering
\caption{\dx{Results on MathQA-Python Across Different Backbones}}
\label{tab:mathqa_python_results}
\begin{tabular}{@{}c@{}}
\resizebox{\textwidth}{!}{
\begin{tabular}{lccc ccc ccc}
\toprule
\multirow{2}{*}{\dx{Methods}} 
& \multicolumn{3}{c}{\dx{GLM-Z1-9B-0414}}
& \multicolumn{3}{c}{\dx{Qwen3-30B-A3B-Thinking-2507}}
& \multicolumn{3}{c}{\dx{Qwen3.5-9B}} \\
\cmidrule(lr){2-4} \cmidrule(lr){5-7} \cmidrule(lr){8-10}
& \dx{pass@1(\%)} & \dx{$\lambda_{\mathrm{avg}}$} & \dx{$\rho$(\%)}
& \dx{pass@1(\%)} & \dx{$\lambda_{\mathrm{avg}}$} & \dx{$\rho$(\%)}
& \dx{pass@1(\%)} & \dx{$\lambda_{\mathrm{avg}}$} & \dx{$\rho$(\%)} \\
\midrule
\dx{Base}      & \dx{78.5} & \dx{2162} & \dx{--}   & \dx{84.2} & \dx{2022} & \dx{--}   & \dx{80.5} & \dx{1149} & \dx{--} \\
\dx{Plain SFT} & \dx{79.8} & \dx{1639} & \dx{24.2} & \dx{84.3} & \dx{1953} & \dx{3.4}  & \dx{82.1} & \dx{1313} & \dx{-14.3} \\
\dx{SEER}      & \dx{\textbf{81.5}} & \dx{\textbf{1210}} & \dx{\textbf{44.0}} & \dx{\textbf{84.3}} & \dx{\textbf{1798}} & \dx{\textbf{11.1}}  & \dx{\textbf{83.1}} & \dx{\textbf{664}}  & \dx{\textbf{42.2}} \\
\bottomrule
\end{tabular}}
\end{tabular}
\begin{tablenotes}[para,flushleft]
    \item[*] \dx{The Base model refers to the corresponding untuned backbone model.}
\end{tablenotes}
\end{threeparttable}
\end{table}

\dx{Table~\ref{tab:mathqa_python_results} shows that \name improves the efficiency--effectiveness trade-off across all three backbones. On GLM-Z1-9B-0414, \name increases pass@1 from 78.5\% to 81.5\% and reduces the average output length by 44.0\%, outperforming Plain SFT in both accuracy and compression. On Qwen3-30B-A3B-Thinking-2507, \name matches Plain SFT in pass@1 (84.3\%) while improving compression from 3.4\% to 11.1\%, showing that it remains useful with LoRA on a larger MoE model. On Qwen3.5-9B, \name reaches the best pass@1 (83.1\%) and a 42.2\% compression rate, whereas Plain SFT makes outputs longer (-14.3\% compression). Overall, these results suggest that \name is not tied to a single model family, architecture, or modality.}

\begin{center}
    \resizebox{\linewidth}{!}{
\begin{tabular}{l!{\vrule width 1pt}p{0.9\columnwidth}}
    \makecell{{\LARGE \faLightbulbO}}  &
    \textbf{Answer to RQ2:}
     \dx{On unseen benchmarks, \name improves accuracy on HumanEval, MBPP, and LiveCodeBench by up to 9.8, 3.2, and 9.2 percentage points, while shortening CoT by up to 40\%.} \dx{On other backbones, \name improves accuracy and compression on \texttt{GLM-Z1-9B-0414} and \texttt{Qwen3.5-9B} and \texttt{Qwen3-30B-A3B-Thinking-2507}. \name also remains effective with both SFT and LoRA, showing robustness across training approaches.}
\end{tabular}}
\end{center}

\subsection{RQ3: The Impact of Each Component}
In this section, we conduct ablation studies to isolate the contribution of each module in \name.
We use MathQA-Python for ablation because it represents one of the most classic and widely studied tasks in software engineering (i.e., code generation).
We vary (1) the BoN sampling size $\mathcal{N}$ while disabling the length filter, and (2) the strictness parameter $\alpha$ in the MAD-based cutoff (which determines $\lambda_c$) while removing BoN. We additionally report the full \name configuration that combines both components. All other training and evaluation settings are kept identical. The results are summarized in Table~\ref{tab:ablation}. 

\begin{table}[ht]
\centering
\caption{The Ablation Results of \name on MathQA-Python Dataset}
\label{tab:ablation}
\resizebox{0.95\textwidth}{!}{
\begin{tabular}{llccc}
\toprule
\textbf{Methods} & \textbf{Settings} & \textbf{Pass@1 (\%)} & \textbf{Avg CoT Len $\lambda_{avg}$ (Tokens)} & \textbf{Compression Rate $\rho$ (\%)} \\
\midrule
\textbf{Base} & - & 63.67 & 1,456 & - \\
\midrule
\textbf{Without Filter} & $\mathcal{N}=1$ & 73.49 & 1,191 & 18.20 \\
\rowcolor{gray!15}
 & $\mathcal{N}=3$ & 74.45 & 1,177 & 19.16 \\
 & $\mathcal{N}=5$ & 74.40 & 1,163 & 20.12 \\
\midrule
\textbf{Without BoN} & $\alpha=2$ & 74.40 & 1,071 & 26.44 \\
\rowcolor{gray!15}
 & $\alpha=1.5$ & 74.19 & 918 & 36.95 \\
 & $\alpha=1$ & 73.50 & 878 & 39.70 \\
\rowcolor{gray!15}
 & $\alpha=0.5$ & 72.76 & 836 & 42.58 \\
\midrule
\hiderowcolors
\textbf{\name} & \begin{tabular}[c]{@{}l@{}}$\mathcal{N}=3$\space\space $\alpha=1$\end{tabular} & 74.88 & 877 & 39.77 \\
\bottomrule
\end{tabular}
}
\end{table}

In the \textbf{Without Filter} setting (Table~\ref{tab:ablation}), we vary the BoN sampling size $\mathcal{N}$. Increasing $\mathcal{N}$ from 1 to 3 improves accuracy from 73.49\% to 74.45\%, while the average CoT length decreases slightly from 1,191 to 1,177 tokens. However, further increasing $\mathcal{N}$ to 5 does not provide additional benefit and slightly reduces accuracy (74.40\%), indicating that the effect of BoN quickly saturates. In this setting, the compression rate remains modest (18.20\% to 20.12\%).

In the \textbf{Without BoN} setting (Table~\ref{tab:ablation}), we vary the strictness parameter $\alpha$ in our MAD-based cutoff $\lambda_{\mathrm{mad}}(\alpha)=\tilde{\lambda}+\alpha\cdot\mathrm{MAD}$. As the filter becomes stricter (from $\alpha=2$ to $\alpha=0.5$), the compression rate increases from 26.44\% to 42.58\%, but accuracy drops from 74.40\% to 72.76\%. This shows a clear trade-off: stronger filtering yields higher compression but can remove useful reasoning information.

We use $\mathcal{N}=3$ and $\alpha=1$ as the default setting in \name.
With both components enabled, \name achieves the best balance, reaching the highest accuracy (74.88\%) with a competitive compression rate (39.77\%).

\begin{center}
    \resizebox{\linewidth}{!}{
\begin{tabular}{l!{\vrule width 1pt}p{0.9\columnwidth}}
    \makecell{{\LARGE \faLightbulbO}}  &\textbf{Answer to RQ3:} BoN alone provides only modest gains (e.g., accuracy increases from 73.49\% to 74.45\% when $\mathcal{N}$ increases from 1 to 5, with compression of 18.20\% to 20.12\%). Making the length filter stricter increases compression (from 26.44\% to 42.58\%) but reduces accuracy (from 74.40\% to 72.76\%). Using both components together achieves the best balance, reaching 74.88\% accuracy with 39.77\% compression.
\end{tabular}}
\end{center}

\subsection{RQ4: Effectiveness in Mitigating Reasoning Loops}
Infinite reasoning loops occur when a model repeatedly generates identical reasoning segments until the output reaches the maximum context length and is truncated. This phenomenon wastes computational resources, inflates latency, and often produces incomplete or invalid final answers. Since such loops can arise unpredictably across models of different sizes, they represent a critical robustness concern. In this section, we investigate whether our approach can mitigate this problem.

We evaluate \name on \dx{the four datasets from RQ1}. For each dataset, we fine-tune a dataset-specific model using our method and compare it against the base model.
Table~\ref{tab:loops} summarizes the results. \name substantially reduces reasoning loops in all cases, e.g., from 85 to 23 on MathQA-Python (72.9\%), from 15 to 1 on Code-Search (93.3\%), from 222 to 7 on Defect-Detection (96.8\%), and \dx{from 2,755 to 1,711 on APPS (37.9\%)}.

\begin{table}[ht]
\centering
\caption{The Effectiveness of \name in Mitigating Truncation and Reasoning Loops}
\label{tab:loops}
\resizebox{\columnwidth}{!}{
\begin{tabular}{lcccccccc}
\toprule
\multirow{2}{*}{} &
\multicolumn{2}{c}{\textbf{MathQA-Python ($N$=1883)}} &
\multicolumn{2}{c}{\textbf{Code-Search ($N$=9604)}} &
\multicolumn{2}{c}{\textbf{Defect-Detection ($N$=2732)}} &
\multicolumn{2}{c}{\dx{\textbf{APPS ($N$=5000)}}} \\ 
\cmidrule(l){2-3} \cmidrule(l){4-5} \cmidrule(l){6-7} \cmidrule(l){8-9}
 & \textbf{Truncation \# (\%)} & \textbf{Loops \# (\%)} &
   \textbf{Truncation \# (\%)} & \textbf{Loops \# (\%)} &
   \textbf{Truncation \# (\%)} & \textbf{Loops \# (\%)} &
   \textbf{Truncation \# (\%)} & \textbf{Loops \# (\%)} \\ 
\midrule
Base & 110 (5.8\%) & 85 (4.5\%) & 18 (0.19\%) & 15 (0.16\%) & 235 (8.6\%) & 222 (8.1\%) & \dx{2826 (56.5\%)} & \dx{2755 (55.1\%)} \\
\rowcolor{gray!15}
\name & \textbf{30 (1.6\%)} & \textbf{23 (1.2\%)} & \textbf{1 (0.01\%)} & \textbf{1 (0.01\%)} &
        \textbf{7 (0.26\%)} & \textbf{7 (0.26\%)} & \dx{\textbf{1905 (38.1\%)}} & \dx{\textbf{1711 (34.2\%)}} \\
\bottomrule
\end{tabular}
}
\end{table}

In our Empirical Findings~\ref{finding:truncation}, we find that most truncations in the base model are directly caused by such loops.
This behavior wastes much of the available context budget and prevents the model from completing its reasoning. Our adaptive filtering strategy reduces CoT length, mitigates loop formation and reduces truncation, as shorter reasoning provides fewer chances for repetitive output. 

\dx{The APPS results further show that loop mitigation becomes more important on complex problems. On APPS, the base model has loops in more than half of the samples (2,755/5,000, 55.1\%), while \name reduces the looping rate to 34.2\% and the truncation rate from 56.5\% to 38.1\%. This helps explain the RQ1 result in Table~\ref{tab:baselines}, where \name improves APPS accuracy from 30.8\% to 37.5\%, about 7 percentage points over the base model.}

Overall, these results confirm that \name offers a robust defense against infinite reasoning loops, complementing its primary goal of efficient CoT compression and providing benefits for latency, reliability, and output quality.

\begin{center}
    \resizebox{\linewidth}{!}{
\begin{tabular}{l!{\vrule width 1pt}p{0.9\columnwidth}}
    \makecell{{\LARGE \faLightbulbO}}  &\textbf{Answer to RQ4:}
    \name effectively suppresses infinite reasoning loops across all datasets, with reductions ranging from \dx{37.9\% on APPS} to 96.8\% on Defect-Detection. Because most truncations in the base model stem from such loops, the adaptive filtering strategy not only prevents looping problems but also lowers truncation frequency. 
\end{tabular}}
\end{center}

\section{Discussion}\label{sec:discussion}

\subsection{Threats to Validity}

\subsubsection{Internal Validity}
Internal validity concerns whether the observed improvements of \name truly arise from the proposed framework rather than confounding factors. One potential threat is the reliance on a single model family, namely \texttt{DeepSeek-R1-Distill-Qwen-7B}. Although all methods are fine-tuned and evaluated under consistent settings, some effects may reflect properties of this architecture. \dx{To reduce this threat, we include a matched \textit{Plain SFT} baseline that uses the same self-generated training data and optimization schedule as \name, but without BoN refinement or adaptive filtering. We also conduct ablation studies by varying BoN sampling and adaptive filtering thresholds. These controls increase confidence that the gains come from the core design of \name rather than fine-tuning alone.}

Another potential concern is filtering bias. Our filter may exclude some difficult instances that naturally require longer CoTs.
This is partially intended: \name does not aim to create reasoning ability from scratch, but to encourage an already reasoning-capable model to produce concise, non-redundant reasoning.
Since extremely long CoTs are often dominated by overthinking or looping, filtering these traces helps the model learn efficient reasoning patterns.
To avoid removing problems entirely, filtering is applied to CoT traces rather than problem instances via BoN sampling.
More importantly, our method does not incur a pass@1 loss after filtering, suggesting that this filtering strategy helps the model learn to generate reasoning that is both concise and effective.
\dx{A related concern is our GPT-4o-based compression analysis in Section~\ref{sec:empirical}. This analysis is only used as an approximate check of redundancy. It does not prove that the retained text is the only correct or sufficient reasoning trace.}

\subsubsection{External Validity}
External validity concerns whether our findings generalize beyond the evaluated settings. A potential limitation is that our study focuses on \dx{one backbone and a limited set of software engineering benchmarks}. \dx{Although we evaluate \name on four software engineering benchmarks, including two code-generation datasets, defect detection, and code search, this evidence should not be viewed as broad cross-model or cross-benchmark generality. More challenging settings, such as LiveCodeBench- or SWE-Bench-style tasks, remain important directions for future work.}

Another concern is sensitivity to prompt design. To reduce this risk, all methods are evaluated using fixed prompts and consistent prompting strategies, ensuring that observed improvements are attributable to the proposed method rather than prompt-specific effects. Moreover, \name is designed to be largely prompt-agnostic, \dx{although more validation with other prompt styles is still needed}.

Finally, the effectiveness of CoT compression may vary with problem difficulty. \dx{Although the evaluated datasets naturally contain problems of varying difficulty, we do not explicitly group instances by complexity, and our HumanEval/129 analysis should be read only as a case study. In addition, this paper uses token length and truncation as efficiency proxies rather than directly measuring wall-clock latency, throughput, or GPU/CPU memory consumption.} A finer-grained analysis by difficulty \dx{and direct system-level resource measurements are} left for future work.

\section{Related Works}\label{sec:related_work}
Since the concept of CoT was first introduced~\cite{wei2022chain}, research on controlling and optimizing CoT length has drawn increasing attention. Numerous studies have shown that extending intermediate reasoning steps within CoT prompting can significantly enhance the capabilities of LLMs~\cite{fu2022complexity,merrill2023expressive,jin2024impact}. 
However, other work~\cite{wu2025more} indicates that performance does not increase indefinitely; instead, it follows a rise-and-fall pattern as CoT length grows. This suggests that \emph{optimally controlling CoT length} is a key challenge for balancing reasoning quality and efficiency.

\noindent $\blacksquare$ \textbf{Prompt-Guided Efficient Reasoning.}
Prompt-based approaches improve reasoning efficiency by explicitly guiding the model to be concise. 
CCoT~\cite{cheng2024compressed} uses a concise CoT template to encourage brevity through prompt wording. 
Token-Budget~\cite{han2024token} proposes a token-budget-aware reasoning framework that dynamically adjusts reasoning length based on estimated complexity. 
CoD~\cite{xu2025chain} limits reasoning to within five words via a ``Chain-of-Draft'' template.  
Token Complexity~\cite{lee2025well} systematically analyzes prompt compression strategies, highlighting that many still operate far from theoretical efficiency limits.

These methods can reduce verbosity without modifying model parameters, but they are inherently sensitive to prompt design and task distribution. In contrast, our method removes this dependency: by learning from filtered, concise, and correct CoT examples, the model internalizes concise reasoning patterns that generalize across prompts and domains.

\noindent $\blacksquare$ \textbf{Model-Optimization Based Efficient Reasoning.}
Other approaches modify the model or its training data to achieve concise reasoning.
C3oT~\cite{kang_c3ot_2025} trains models jointly on long and short CoTs to learn mappings between them, using conditioned inference to produce short CoTs.  
TokenSkip~\cite{xia_tokenskip_2025} enables skipping less important tokens during generation, allowing controllable compression but sometimes causing instability (e.g., reasoning loops in our observations).  
LightThinker~\cite{zhang_lightthinker_2025} compresses verbose thoughts into compact internal representations, discarding the original CoT from the context.  
CODI~\cite{shen_codi_2025} distills CoT into a continuous space, jointly learning explicit and implicit reasoning traces.

While these methods reduce reasoning length, many require additional compression modules, complex training objectives, or implicit representations that limit interpretability. Our method is simpler: it uses the model's own generated data, combined with a lightweight Best-of-N sampling and adaptive filtering, to directly fine-tune for concise yet explicit reasoning. This avoids extra components and preserves the transparency of reasoning steps.

\noindent $\blacksquare$ \textbf{Chain-of-Thought Effectiveness in Software Engineering.}
CoT has shown strong benefits in software engineering tasks. SCoT~\cite{li2025structured} was among the first to apply structured CoT to code generation. The emergence of DeepSeek-R1~\cite{deepseekai2025deepseekr1} further highlighted the value of streamlined CoT for software engineering tasks.  
Liu et al.~\cite{liu2025revisiting} revisited CoT training paradigms, showing that some seemingly effective methods do not transfer well to real-world coding tasks. 
UnCert-CoT~\cite{zhu2025uncertainty} uses uncertainty estimation to decide when to trigger CoT, avoiding unnecessary reasoning.  
CoD~\cite{yang2025chain} adapts the ``Chain-of-Draft'' approach to software engineering tasks, greatly reducing token usage.

However, existing methods often target specific task scenarios or rely on inference-time control, which may fail to eliminate severe failure modes such as looping behaviors. Our method differs by explicitly filtering out redundant or looping reasoning during training, allowing the model to internalize concise, stable reasoning behaviors. This makes it robust in software engineering tasks while retaining interpretability and avoiding the need for task-specific prompt engineering.

\section{Conclusion}\label{sec:conclusion}

In this paper, we show that overly long CoTs in software engineering tasks increase token cost and truncation risk, with many truncations associated with reasoning loops. We propose \name, which combines BoN sampling and adaptive length filtering to learn concise reasoning from self-generated outputs. 
Across four benchmarks on the evaluated backbone, SEER reduces CoT length by 34.6\% on average while improving performance and generalizing to unseen benchmarks. Future work will evaluate larger models and harder tasks and directly measure latency and memory costs.

\section{Data Availability}
Data supporting this research is available at the following link: \href{https://github.com/huangkr03/SEER}{\name}.

\begin{acks}
This work was supported by the National Key R\&D Program of China (No. 2024YFB4506400) and sponsored by CCF-Huawei Populus Grove Fund.
\end{acks}

\bibliographystyle{ACM-Reference-Format}

\bibliography{ref}

@article{wei2022chain,
  title={Chain-of-thought prompting elicits reasoning in large language models},
  author={Wei, Jason and Wang, Xuezhi and Schuurmans, Dale and Bosma, Maarten and Xia, Fei and Chi, Ed and Le, Quoc V and Zhou, Denny and others},
  journal={Advances in neural information processing systems},
  volume={35},
  pages={24824--24837},
  year={2022}
,
  doi={10.48550/arXiv.2201.11903},
}

@article{NEURIPS2022_8bb0d291,
  title={Large language models are zero-shot reasoners},
  author={Kojima, Takeshi and Gu, Shixiang Shane and Reid, Machel and Matsuo, Yutaka and Iwasawa, Yusuke},
  journal={Advances in neural information processing systems},
  volume={35},
  pages={22199--22213},
  year={2022}
,
  doi={10.48550/arXiv.2205.11916},
}

@article{nye2021workscratchpadsintermediatecomputation,
  title={Show Your Work: Scratchpads for Intermediate Computation with Language Models},
  author={Nye, Maxwell and Andreassen, Anders Johan and Gur-Ari, Guy and Michalewski, Henryk and Austin, Jacob and Bieber, David and Dohan, David and Lewkowycz, Aitor and Bosma, Maarten and Luan, David and others},
  journal={arXiv preprint arXiv:2112.00114},
  year={2021}
,
  doi={10.48550/arXiv.2112.00114},
}

@inproceedings{wang2023selfconsistency,
title={Self-Consistency Improves Chain of Thought Reasoning in Language Models},
author={Xuezhi Wang and Jason Wei and Dale Schuurmans and Quoc V Le and Ed H. Chi and Sharan Narang and Aakanksha Chowdhery and Denny Zhou},
booktitle={The Eleventh International Conference on Learning Representations },
year={2023},
  doi={10.48550/arXiv.2203.11171},
url={https://openreview.net/forum?id=1PL1NIMMrw}
}

@article{zhou2023leasttomost,
  title={Least-to-most prompting enables complex reasoning in large language models},
  author={Zhou, Denny and Sch{\"a}rli, Nathanael and Hou, Le and Wei, Jason and Scales, Nathan and Wang, Xuezhi and Schuurmans, Dale and Cui, Claire and Bousquet, Olivier and Le, Quoc and others},
  journal={arXiv preprint arXiv:2205.10625},
  year={2022}
,
  doi={10.48550/arXiv.2205.10625},
}

@inproceedings{agrawal2024taming,
  title={Taming $\{$Throughput-Latency$\}$ tradeoff in $\{$LLM$\}$ inference with $\{$Sarathi-Serve$\}$},
  author={Agrawal, Amey and Kedia, Nitin and Panwar, Ashish and Mohan, Jayashree and Kwatra, Nipun and Gulavani, Bhargav and Tumanov, Alexey and Ramjee, Ramachandran},
  booktitle={18th USENIX Symposium on Operating Systems Design and Implementation (OSDI 24)},
  pages={117--134},
  year={2024}
,
  doi={10.48550/arXiv.2403.02310},
}

@article{ho2024advances,
  title={Block transformer: Global-to-local language modeling for fast inference},
  author={Ho, Namgyu and Bae, Sangmin and Kim, Taehyeon and Jo, Hyunjik and Kim, Yireun and Schuster, Tal and Fisch, Adam and Thorne, James and Yun, Se-Young},
  journal={Advances in Neural Information Processing Systems},
  volume={37},
  pages={48740--48783},
  year={2024}
,
  doi={10.48550/arXiv.2406.02657},
}

@misc{xia_tokenskip_2025,
	title = {{TokenSkip}: {Controllable} {Chain}-of-{Thought} {Compression} in {LLMs}},
	copyright = {arXiv.org perpetual, non-exclusive license},
	shorttitle = {{TokenSkip}},
	url = {https://arxiv.org/abs/2502.12067},
	doi = {10.48550/ARXIV.2502.12067},
	urldate = {2025-03-18},
	publisher = {arXiv},
	author = {Xia, Heming and Li, Yongqi and Leong, Chak Tou and Wang, Wenjie and Li, Wenjie},
	year = {2025},
	note = {Version Number: 1}
}

@misc{yu_distilling_2024,
	title = {Distilling {System} 2 into {System} 1},
	copyright = {Creative Commons Attribution Non Commercial No Derivatives 4.0 International},
	url = {https://arxiv.org/abs/2407.06023},
	doi = {10.48550/ARXIV.2407.06023},
	urldate = {2025-03-29},
	publisher = {arXiv},
	author = {Yu, Ping and Xu, Jing and Weston, Jason and Kulikov, Ilia},
	year = {2024},
	note = {Version Number: 3}
}

@misc{zhang_lightthinker_2025,
	title = {{LightThinker}: {Thinking} {Step}-by-{Step} {Compression}},
	copyright = {arXiv.org perpetual, non-exclusive license},
	shorttitle = {{LightThinker}},
	url = {https://arxiv.org/abs/2502.15589},
	doi = {10.48550/ARXIV.2502.15589},
	urldate = {2025-03-29},
	publisher = {arXiv},
	author = {Zhang, Jintian and Zhu, Yuqi and Sun, Mengshu and Luo, Yujie and Qiao, Shuofei and Du, Lun and Zheng, Da and Chen, Huajun and Zhang, Ningyu},
	year = {2025},
	note = {Version Number: 1}
}

@misc{shen_codi_2025,
	title = {{CODI}: {Compressing} {Chain}-of-{Thought} into {Continuous} {Space} via {Self}-{Distillation}},
	copyright = {Creative Commons Attribution 4.0 International},
	shorttitle = {{CODI}},
	url = {https://arxiv.org/abs/2502.21074},
	doi = {10.48550/ARXIV.2502.21074},
	urldate = {2025-04-09},
	publisher = {arXiv},
	author = {Shen, Zhenyi and Yan, Hanqi and Zhang, Linhai and Hu, Zhanghao and Du, Yali and He, Yulan},
	year = {2025},
	note = {Version Number: 1},
}

@misc{chen2021evaluating,
      title={Evaluating Large Language Models Trained on Code},
      author={Chen et al.},
      year={2021},
  doi={10.48550/arXiv.2107.03374},
      eprint={2107.03374},
      archivePrefix={arXiv},
      primaryClass={cs.LG}
}

@article{austin2021program,
  title={Program Synthesis with Large Language Models},
  author={Austin, Jacob and Odena, Augustus and Nye, Maxwell and Bosma, Maarten and Michalewski, Henryk and Dohan, David and Jiang, Ellen and Cai, Carrie and Terry, Michael and Le, Quoc and others},
  journal={arXiv preprint arXiv:2108.07732},
  year={2021}
,
  doi={10.48550/arXiv.2108.07732},
}

@article{kang_c3ot_2025,
	title = {{C3oT}: {Generating} {Shorter} {Chain}-of-{Thought} {Without} {Compromising} {Effectiveness}},
	volume = {39},
	issn = {2374-3468, 2159-5399},
	shorttitle = {{C3oT}},
	url = {https://ojs.aaai.org/index.php/AAAI/article/view/34608},
	doi = {10.1609/aaai.v39i23.34608},
	number = {23},
	urldate = {2025-04-14},
	journal = {Proceedings of the AAAI Conference on Artificial Intelligence},
	author = {Kang, Yu and Sun, Xianghui and Chen, Liangyu and Zou, Wei},
	month = apr,
	year = {2025},
	pages = {24312--24320},
}

@misc{deepseekai2025deepseekr1,
      title={DeepSeek-R1: Incentivizing Reasoning Capability in LLMs via Reinforcement Learning}, 
      author={DeepSeek-AI and Guo, Daya and Yang, Dejian and Zhang, Haowei and Song, Junxiao and Wang, Peiyi and Zhu, Qihao and Xu, Runxin and Zhang, Ruoyu and Ma, Shirong and others},
      year={2025},
  doi={10.48550/arXiv.2501.12948},
      eprint={2501.12948},
      archivePrefix={arXiv},
      primaryClass={cs.CL},
      url={https://arxiv.org/abs/2501.12948}, 
}

@article{fu2022complexity,
  title={Complexity-based prompting for multi-step reasoning},
  author={Fu, Yao and Peng, Hao and Sabharwal, Ashish and Clark, Peter and Khot, Tushar},
  journal={arXiv preprint arXiv:2210.00720},
  year={2022}
,
  doi={10.48550/arXiv.2210.00720},
}

@article{merrill2023expressive,
  title={The expressive power of transformers with chain of thought},
  author={Merrill, William and Sabharwal, Ashish},
  journal={arXiv preprint arXiv:2310.07923},
  year={2023}
,
  doi={10.48550/arXiv.2310.07923},
}

@article{jin2024impact,
  title={The impact of reasoning step length on large language models},
  author={Jin, Mingyu and Yu, Qinkai and Shu, Dong and Zhao, Haiyan and Hua, Wenyue and Meng, Yanda and Zhang, Yongfeng and Du, Mengnan},
  journal={arXiv preprint arXiv:2401.04925},
  year={2024}
,
  doi={10.48550/arXiv.2401.04925},
}

@article{wu2025more,
  title={When More is Less: Understanding Chain-of-Thought Length in LLMs},
  author={Wu, Yuyang and Wang, Yifei and Ye, Ziyu and Du, Tianqi and Jegelka, Stefanie and Wang, Yisen},
  journal={arXiv preprint arXiv:2502.07266},
  year={2025}
,
  doi={10.48550/arXiv.2502.07266},
}

@article{cheng2024compressed,
  title={Compressed chain of thought: Efficient reasoning through dense representations},
  author={Cheng, Jeffrey and Van Durme, Benjamin},
  journal={arXiv preprint arXiv:2412.13171},
  year={2024}
,
  doi={10.48550/arXiv.2412.13171},
}

@article{han2024token,
  title={Token-budget-aware llm reasoning},
  author={Han, Tingxu and Wang, Zhenting and Fang, Chunrong and Zhao, Shiyu and Ma, Shiqing and Chen, Zhenyu},
  journal={arXiv preprint arXiv:2412.18547},
  year={2024}
,
  doi={10.48550/arXiv.2412.18547},
}

@article{xu2025chain,
  title={Chain of draft: Thinking faster by writing less},
  author={Xu, Silei and Xie, Wenhao and Zhao, Lingxiao and He, Pengcheng},
  journal={arXiv preprint arXiv:2502.18600},
  year={2025}
,
  doi={10.48550/arXiv.2502.18600},
}

@article{lee2025well,
  title={How well do llms compress their own chain-of-thought? a token complexity approach},
  author={Lee, Ayeong and Che, Ethan and Peng, Tianyi},
  journal={arXiv preprint arXiv:2503.01141},
  year={2025}
,
  doi={10.48550/arXiv.2503.01141},
}

@article{wang2025think,
  title={Think or Not? Selective Reasoning via Reinforcement Learning for Vision-Language Models},
  author={Wang, Jiaqi and Lin, Kevin Qinghong and Cheng, James and Shou, Mike Zheng},
  journal={arXiv preprint arXiv:2505.16854},
  year={2025}
,
  doi={10.48550/arXiv.2505.16854},
}

@article{chen2024not,
  title={Do not think that much for 2+ 3=? on the overthinking of o1-like llms},
  author={Chen, Xingyu and Xu, Jiahao and Liang, Tian and He, Zhiwei and Pang, Jianhui and Yu, Dian and Song, Linfeng and Liu, Qiuzhi and Zhou, Mengfei and Zhang, Zhuosheng and others},
  journal={arXiv preprint arXiv:2412.21187},
  year={2024}
,
  doi={10.48550/arXiv.2412.21187},
}

@article{xie2025logic,
  title={Logic-rl: Unleashing llm reasoning with rule-based reinforcement learning},
  author={Xie, Tian and Gao, Zitian and Ren, Qingnan and Luo, Haoming and Hong, Yuqian and Dai, Bryan and Zhou, Joey and Qiu, Kai and Wu, Zhirong and Luo, Chong},
  journal={arXiv preprint arXiv:2502.14768},
  year={2025}
,
  doi={10.48550/arXiv.2502.14768},
}

@article{cuadron2025danger,
  title={The Danger of Overthinking: Examining the Reasoning-Action Dilemma in Agentic Tasks},
  author={Cuadron, Alejandro and Li, Dacheng and Ma, Wenjie and Wang, Xingyao and Wang, Yichuan and Zhuang, Siyuan and Liu, Shu and Schroeder, Luis Gaspar and Xia, Tian and Mao, Huanzhi and others},
  journal={arXiv preprint arXiv:2502.08235},
  year={2025}
,
  doi={10.48550/arXiv.2502.08235},
}

@article{chen2024unlocking,
  title={Unlocking the capabilities of thought: A reasoning boundary framework to quantify and optimize chain-of-thought},
  author={Chen, Qiguang and Qin, Libo and Wang, Jiaqi and Zhou, Jingxuan and Che, Wanxiang},
  journal={Advances in Neural Information Processing Systems},
  volume={37},
  pages={54872--54904},
  year={2024}
,
  doi={10.48550/arXiv.2410.05695},
}

@misc{munkhbat2025selftraining,
	title = {Self-{Training} {Elicits} {Concise} {Reasoning} in {Large} {Language} {Models}},
	copyright = {Creative Commons Attribution 4.0 International},
	url = {https://arxiv.org/abs/2502.20122},
	doi = {10.48550/ARXIV.2502.20122},
	urldate = {2025-03-29},
	publisher = {arXiv},
	author = {Munkhbat, Tergel and Ho, Namgyu and Kim, Seo Hyun and Yang, Yongjin and Kim, Yujin and Yun, Se-Young},
	year = {2025},
	note = {Version Number: 2},
}

@article{xu2025mind,
  title={Mind the Gap: Bridging Thought Leap for Improved Chain-of-Thought Tuning},
  author={Xu, Haolei and Yan, Yuchen and Shen, Yongliang and Zhang, Wenqi and Hou, Guiyang and Jiang, Shengpei and Song, Kaitao and Lu, Weiming and Xiao, Jun and Zhuang, Yueting},
  journal={arXiv preprint arXiv:2505.14684},
  year={2025}
,
  doi={10.48550/arXiv.2505.14684},
}

@article{zhuang2025accelerating,
  title={Accelerating chain-of-thought reasoning: When goal-gradient importance meets dynamic skipping},
  author={Zhuang, Ren and Wang, Ben and Sun, Shuifa},
  journal={arXiv preprint arXiv:2505.08392},
  year={2025}
,
  doi={10.48550/arXiv.2505.08392},
}

@article{li2025structured,
  title={Structured chain-of-thought prompting for code generation},
  author={Li, Jia and Li, Ge and Li, Yongmin and Jin, Zhi},
  journal={ACM Transactions on Software Engineering and Methodology},
  volume={34},
  number={2},
  pages={1--23},
  year={2025},
  doi={10.1145/3690635},
  publisher={ACM New York, NY}
}

@article{yao2023tree,
  title={Tree of thoughts: Deliberate problem solving with large language models, 2023},
  author={Yao, Shunyu and Yu, Dian and Zhao, Jeffrey and Shafran, Izhak and Griffiths, Thomas L and Cao, Yuan and Narasimhan, Karthik},
  journal={URL https://arxiv. org/abs/2305.10601},
  volume={3},
  pages={1},
  year={2023}
,
  doi={10.48550/arXiv.2305.10601},
}

@article{Lu2021CodeXGLUEAM,
  title={CodeXGLUE: A Machine Learning Benchmark Dataset for Code Understanding and Generation},
  author={Shuai Lu and Daya Guo and Shuo Ren and Junjie Huang and Alexey Svyatkovskiy and Ambrosio Blanco and Colin Clement and Dawn Drain and Daxin Jiang and Duyu Tang and Ge Li and Lidong Zhou and Linjun Shou and Long Zhou and Michele Tufano and Ming Gong and Ming Zhou and Nan Duan and Neel Sundaresan and Shao Kun Deng and Shengyu Fu and Shujie Liu},
  journal={ArXiv},
  year={2021},
  doi={10.48550/arXiv.2102.04664},
  volume={abs/2102.04664}
}

@article{husain2019codesearchnet,
  title={Codesearchnet challenge: Evaluating the state of semantic code search},
  author={Husain, Hamel and Wu, Ho-Hsiang and Gazit, Tiferet and Allamanis, Miltiadis and Brockschmidt, Marc},
  journal={arXiv preprint arXiv:1909.09436},
  year={2019}
,
  doi={10.48550/arXiv.1909.09436},
}

@inproceedings{zhou2019devign,
  title={Devign: Effective vulnerability identification by learning comprehensive program semantics via graph neural networks},
  author={Zhou, Yaqin and Liu, Shangqing and Siow, Jingkai and Du, Xiaoning and Liu, Yang},
  booktitle={Advances in Neural Information Processing Systems},
  pages={10197--10207},
  year={2019}
,
  doi={10.48550/arXiv.1909.03496},
}

@article{hendrycksapps2021,
  title={Measuring Coding Challenge Competence With APPS},
  author={Dan Hendrycks and Steven Basart and Saurav Kadavath and Mantas Mazeika and Akul Arora and Ethan Guo and Collin Burns and Samir Puranik and Horace He and Dawn Song and Jacob Steinhardt},
  journal={NeurIPS},
  year={2021}
,
  doi={10.48550/arXiv.2105.09938},
}

@inproceedings{liu2025revisiting,
  title={Revisiting Chain-of-Thought in Code Generation: Do Language Models Need to Learn Reasoning before Coding?},
  author={Liu, Ren-Biao and Li, Anqi and Yang, Chaoding and Sun, Hui and Li, Ming},
  year={2025},
  booktitle={Proceedings of the 42nd International Conference on Machine Learning},
  pages={38809--38826},
  doi={10.5555/3780338.3781887}
}

@article{zhu2025uncertainty,
  title={Uncertainty-guided chain-of-thought for code generation with llms},
  author={Zhu, Yuqi and Li, Ge and Jiang, Xue and Li, Jia and Mei, Hong and Jin, Zhi and Dong, Yihong},
  journal={arXiv preprint arXiv:2503.15341},
  year={2025},
  doi={10.48550/arXiv.2503.15341},
}

@article{yang2025chain,
  title={Chain of Draft for Software Engineering: Challenges in Applying Concise Reasoning to Code Tasks},
  author={Yang, Shaoyi},
  journal={arXiv preprint arXiv:2506.10987},
  year={2025}
,
  doi={10.48550/arXiv.2506.10987},
}

@article{zheng2024llamafactory,
  title={Llamafactory: Unified efficient fine-tuning of 100+ language models},
  author={Zheng, Yaowei and Zhang, Richong and Zhang, Junhao and Ye, Yanhan and Luo, Zheyan and Feng, Zhangchi and Ma, Yongqiang},
  journal={arXiv preprint arXiv:2403.13372},
  year={2024}
,
  doi={10.48550/arXiv.2403.13372},
}

@misc{autogpt_recursive_loop,
  title        = {User Feedback leads to endless thinking freeze and not writing},
  author       = {{Significant Gravitas}},
  year         = {2023},
  howpublished = {\url{https://github.com/Significant-Gravitas/Auto-GPT/issues/1591}},
  note         = {GitHub Issue \#1591}
}

@misc{autogpt_do_nothing,
  title        = {Command do\_nothing returned: No action performed},
  author       = {{Significant Gravitas}},
  year         = {2023},
  howpublished = {\url{https://github.com/Significant-Gravitas/Auto-GPT/issues/830}},
  note         = {GitHub Issue \#830}
}

@misc{langgraph_repetition,
  title        = {Getting double repetitive output from agents tools},
  author       = {{LangChain AI}},
  year         = {2024},
  howpublished = {\url{https://github.com/langchain-ai/langgraph/issues/3062}},
  note         = {GitHub Issue \#3062}
}

@misc{pipis2025reasoningloop,
      title={Wait, Wait, Wait... Why Do Reasoning Models Loop?}, 
      author={Charilaos Pipis and Shivam Garg and Vasilis Kontonis and Vaishnavi Shrivastava and Akshay Krishnamurthy and Dimitris Papailiopoulos},
      year={2025},
  doi={10.48550/arXiv.2512.12895},
      eprint={2512.12895},
      archivePrefix={arXiv},
      primaryClass={cs.LG},
      url={https://arxiv.org/abs/2512.12895}, 
}

@misc{gpt4o,
  title       = {Hello GPT-4o},
  author      = {{OpenAI}},
  year        = {2024},
  url         = {https://openai.com/index/hello-gpt-4o/},
}

@inproceedings{NEURIPS2023_91f18a12,
 author = {Zheng, Lianmin and Chiang, Wei-Lin and Sheng, Ying and Zhuang, Siyuan and Wu, Zhanghao and Zhuang, Yonghao and Lin, Zi and Li, Zhuohan and Li, Dacheng and Xing, Eric and Zhang, Hao and Gonzalez, Joseph E and Stoica, Ion},
 booktitle = {Advances in Neural Information Processing Systems},
 editor = {A. Oh and T. Naumann and A. Globerson and K. Saenko and M. Hardt and S. Levine},
 pages = {46595--46623},
 publisher = {Curran Associates, Inc.},
 title = {Judging LLM-as-a-Judge with MT-Bench and Chatbot Arena},
 url = {https://proceedings.neurips.cc/paper_files/paper/2023/file/91f18a1287b398d378ef22505bf41832-Paper-Datasets_and_Benchmarks.pdf},
 volume = {36},
 year = {2023}
,
  doi={10.48550/arXiv.2306.05685},
}

@misc{qwen3technicalreport,
      title={Qwen3 Technical Report}, 
      author={Yang, An and Li, Anfeng and Yang, Baosong and Zhang, Beichen and Hui, Binyuan and Zheng, Bo and Yu, Bowen and Gao, Chang and Huang, Chengen and Lv, Chenxu and others},
      year={2025},
  doi={10.48550/arXiv.2505.09388},
      eprint={2505.09388},
      archivePrefix={arXiv},
      primaryClass={cs.CL},
      url={https://arxiv.org/abs/2505.09388}, 
}

\end{document}